\documentclass[a4paper,11pt]{article}
\pdfoutput=1
\usepackage{jheppub}
\usepackage{siunitx}
\usepackage[svgnames]{xcolor}
\usepackage[utf8]{inputenc}

\usepackage[textwidth=2cm,			% Breite der Todo-Einträge
textsize=footnotesize,	% Schriftgröße der Einträge
german,					% deutsche Beschriftungen
shadow,					% Schlagschatten für Boxen (weils so hübsch ist)
colorinlistoftodos]{todonotes}% farbige Markierungen für unterschiedliche Aufgabentypen

\preprint{MS-TP-20-31}

\title{Impact of scale, nuclear PDF and temperature variations on the interpretation of medium-modified jet production data from the LHC}

\author[a]{A. Andronic,}
\author[a]{J. Honermann,}
\author[b]{M. Klasen,}
\author[a]{C. Klein-B{\"o}sing,}
\author[b]{and J. Salomon}

\affiliation[a]{Institut f{\"u}r Kernphysik,\\Wilhelm-Klemm-Str. 9, 48149 M{\"u}nster, Germany}
\affiliation[b]{Institut f{\"u}r Theoretische Physik,\\Wilhelm-Klemm-Str. 9, 48149 M{\"u}nster, Germany}

\emailAdd{andronic@uni-muenster.de}
\emailAdd{jan.honermann@uni-muenster.de}
\emailAdd{michael.klasen@uni-muenster.de}
\emailAdd{christian.klein-boesing@uni-muenster.de}
\emailAdd{jens.salomon@uni-muenster.de} 

\abstract{In this paper we present a study of in-medium jet modifications performed with JEWEL and PYTHIA 6.4, focusing on the uncertainties related to variations of the perturbative scales and nuclear parton distribution functions (PDFs) and on the impact of the initial and crossover temperature variations of the medium. The simulations are compared to LHC data for the jet spectrum and the nuclear modification factor. We assess the interplay between the choice of nuclear PDFs and different medium parameters and study the impact of nuclear PDFs and the medium on the jet structure via the Lund plane.}

\begin{document} 

\maketitle

\section{Introduction}
\label{sec:1}

At sufficiently large temperatures, a phase transition is expected in strongly interacting matter: from a phase, where quarks and gluons are confined into hadrons, to a quark-gluon plasma (QGP), where they can move over distances much larger than the size of a hadron. This transition can be calculated in discrete formulations of Quantum Chromodynamics (lattice QCD) at vanishing baryon density, where a crossover transition at $T_{\rm C} \approx 156$~MeV is found \cite{Borsanyi:2013bia,Bazavov:2018mes}. This pseudo-critical temperature can be confirmed experimentally e.g. in statistical analyses of hadron abundances as the freeze-out temperature occurring in heavy-ion collisions at the Large Hadron Collider (LHC) \cite{Andronic:2018cr}.

For a detailed study of the hot QGP phase, well calibrated probes such as parton scatterings prior to the QGP formation are the prime tool. Provided a hard scale (e.g.\ a large momentum transfer or mass), these processes are calculable in perturbative QCD and can be benchmarked by measurements in pp collisions, including the role of non-perturbative hadronisation. In the medium, the scattered partons and the subsequent parton shower experience the full QGP evolution and are modified by the large density of colour charges via elastic collisions and induced gluon radiation.  The corresponding experimental observables sensitive to the hard scattered partons are provided e.g.\ by the reconstruction of jets, the modification of their differential production cross section and their angular structure. 

Since the experimental discovery of jet quenching \cite{Adcox:2001jp} in single (leading) particle production at high transverse momentum ($p_\mathrm{T}$), the nuclear modification factor ($R_{\mathrm{AA}}$) is the most popular observable to quantify medium effects on the particle (jet) production in heavy-ion collisions. It is defined as
\begin{equation}
    R_{\mathrm{AA}} =  \frac{\frac{1}{N_\mathrm{Evt}} \frac{\mathrm{d}^2N_\mathrm{jets}^\mathrm{AA}}{\mathrm{d}p_\mathrm{T} \mathrm{d}\eta}\bigg|_\mathrm{centrality}}{\langle T_\mathrm{AA} \rangle \frac{\mathrm{d}^2\sigma_\mathrm{jets}}{\mathrm{d} p_\mathrm{T}\mathrm{d}\eta}\bigg|_\mathrm{pp}} ,
    \label{eq:RAA}
\end{equation}
%mkl where the differential jet yield $N_\mathrm{jets}^\mathrm{AA}$ in AA collisions for a given centrality range is compared to the equivalent measurement in pp collisions, the differential jet cross section.
%mkl++
where the differential jet yield $N_\mathrm{jets}^\mathrm{AA}$ is the number of jets recorded in $N_{\rm Evt}$ heavy ion collisions with a given centrality. Thus the differential yield in AA is compared to the equivalent measurement in pp collisions, the differential jet cross section.
%mkl--
The latter is scaled by the nuclear thickness function $T_\mathrm{AA}$ to account for the geometric increase of the parton flux in AA collisions
%mkl++
with the given centrality selection.
%mkl--
Its value is determined in a geometric Glauber model~\cite{Miller:2007ri}. In the absence of any medium effects the ratio should be unity in the region where hard processes dominate. At the LHC, a strong suppression of jet production is seen in central PbPb collisions \cite{Adam:2015ewa,Khachatryan:2016jfl,Aaboud:2018twu} compared to pp reactions \cite{Abelev:2013fn}. The measurements of electromagnetic and electroweak probes in PbPb \cite{Chatrchyan:2011ua,Chatrchyan:2012nt,Acharya:2017wpf,Aad:2019sfe,Aad:2019lan} and jets in pPb collisions  \cite{Abelev:2014dsa,Adam:2015hoa,Khachatryan:2016xdg} confirm that the suppression is due to a strong final state effect and not due to a modified initial state in the Pb nucleus. 

For a deeper understanding of the parton shower evolution in the medium, on the one hand more differential observables than the nuclear modification factor are needed. On the other hand, for a precise comparison to predictions one also has to take into account the influence of a modified initial state (i.e. nuclear PDFs) on the production of partonic probes and their strong final state interactions. While the partonic production can be well calibrated in pp collisions, the final state interactions of a scattered parton with a medium and the medium modified parton shower are more challenging due to the involved time and momentum scales and the \mbox{(hydro-)}dynamics of the QGP evolution (for a recent review see \cite{Cao:2020wlm}). At the same time, the modification of the parton densities in the nucleus relevant at low transverse momentum (corresponding to small Bjorken $x$) is affected by a large uncertainty. Thus, the extraction of the QGP-related parameters and effects in the final state necessitates the evaluation of the uncertainty induced by the initial state parton densities and hard scattering alone.

In this paper we study the medium modification of jets with the JEWEL model \cite{Zapp:2008gi,Zapp:2012ak,Zapp:2013vla,KunnawalkamElayavalli:2017hxo} and compare it to measurements of jet production in pp and PbPb collisions at the LHC. We evaluate the effect of nuclear PDFs and their uncertainties and compare it to the impact of scale uncertainties and to the crossover and initial temperature variations in the 1D hydrodynamic model of JEWEL. In addition, the impact of nuclear PDFs and the hot medium on the observed jet structure is studied in JEWEL via the so-called Lund plane \cite{Dreyer:2018nbf}, that maps the occurrence of splittings in angular and momentum space within jets. 

The remainder of this paper is organised as follows: In Sec.\ \ref{sec:2} we describe our theoretical setup, in particular the main features of JEWEL and our choices of nuclear PDFs, perturbative scales, medium geometry and evolution.  Sec.\ \ref{sec:3} contains a description of our data selection and our main results on jet spectra and nuclear modification factors, detailed studies of the dependence of these results on the medium parameters and a study of the jet substructure in vacuum and in the medium using the Lund plane. Our conclusions and an outlook are then given in Sec.\ \ref{sec:4}.

\section{Jet production in JEWEL}
\label{sec:2}

JEWEL \cite{Zapp:2008gi,Zapp:2012ak,Zapp:2013vla,KunnawalkamElayavalli:2017hxo} is a leading order (LO) Monte Carlo event generator that uses PYTHIA 6.4 \cite{Sjostrand:2006za} to generate hard scattering events, parton showers and hadronisation, but in addition supplements the parton showers with medium effects. It is based consistently on perturbative language to describe jet evolution and interactions in the medium in a common framework. By construction, limitations of analytic approaches such as kinematical limitations, momentum conservation, restriction to single gluon emission etc.\ are overcome. Infrared divergences of the cross section are taken into account with a temperature-dependent regulator $\mu_{\rm D}=0.9\cdot 3T$ \cite{Zapp:2013zya}. The Glauber model is used internally to calculate geometry-driven quantities for heavy-ion collisions. The medium evolution in JEWEL is realised through a Bjorken model~\cite{Bjorken:1982qr}, a one-dimensional expansion model. Partons from the initial hard process can interact with the medium, when passing through it. 

\subsection{Initial conditions and hard processes}

\subsubsection{Nuclear PDFs}

In the default JEWEL configuration, the jet production matrix elements and initial parton showers are simulated using central EPS09 LO nuclear PDFs (nPDFs) \cite{Eskola:2009uj}, based on CTEQ6L1 free proton PDFs \cite{Pumplin:2002vw} and provided through the LHAPDF5 interface \cite{Whalley:2005nh}. More recent nPDF sets are available only as part of LHAPDF6 \cite{Buckley:2014ana}. We implement them in our calculations using the interface LHAGLUE for legacy code built around LHAPDF5.

As our default, we employ the full nuclear nCTEQ15 NLO \cite{Kovarik:2015cma} set for lead, together with the free proton PDFs from nCTEQ15. The full nuclear PDFs are constructed as the PDFs of an average nucleon, so using them as \emph{proton} PDFs in JEWEL and turning off the modification factor from EPS09 provides the correct initial state for hard PbPb collisions.\footnote{JEWEL differentiates between proton and neutron PDFs, constructing the latter via isospin symmetry, as well as between pp, pn and nn collision events. Since the nCTEQ15 PDFs describe an average nucleon, it is sufficient to consider only the pp channel.} For consistency with the nCTEQ15 parametrisation, the strong coupling constant $\alpha_s$ is evaluated at next-to-leading order (NLO) with five active flavours and $\Lambda_{\rm QCD}=226$ MeV. The PDF uncertainties are calculated with the $32$ error PDFs of nCTEQ15.

As an alternative, we also use the EPPS16 NLO set~\cite{Eskola:2016oht} based on CT14NLO free proton nPDFs \cite{Dulat:2015mca}. The corresponding value of $\Lambda_{\rm QCD}=208$ MeV can be obtained e.g.\ with RunDec \cite{Chetyrkin:2000yt}.\footnote{Technically, the EPPS16 LHAPDF info file is supplemented with an entry {\tt AlphaS\_Lambda5} corresponding to this value, and the flag {\tt Particle} is modified from the PDG-ID $1000822080$ for Pb to $2212$ for protons.} The PDF uncertainties are calculated with the 40 error PDFs of EPPS16. Uncertainties in the underlying proton PDFs are not taken into account, as they appear in $R_{\rm AA}$ in both the numerator and the denominator and should thus cancel to a large extent.

\begin{figure} %[htb]
	\centering
	\includegraphics[width=\textwidth]{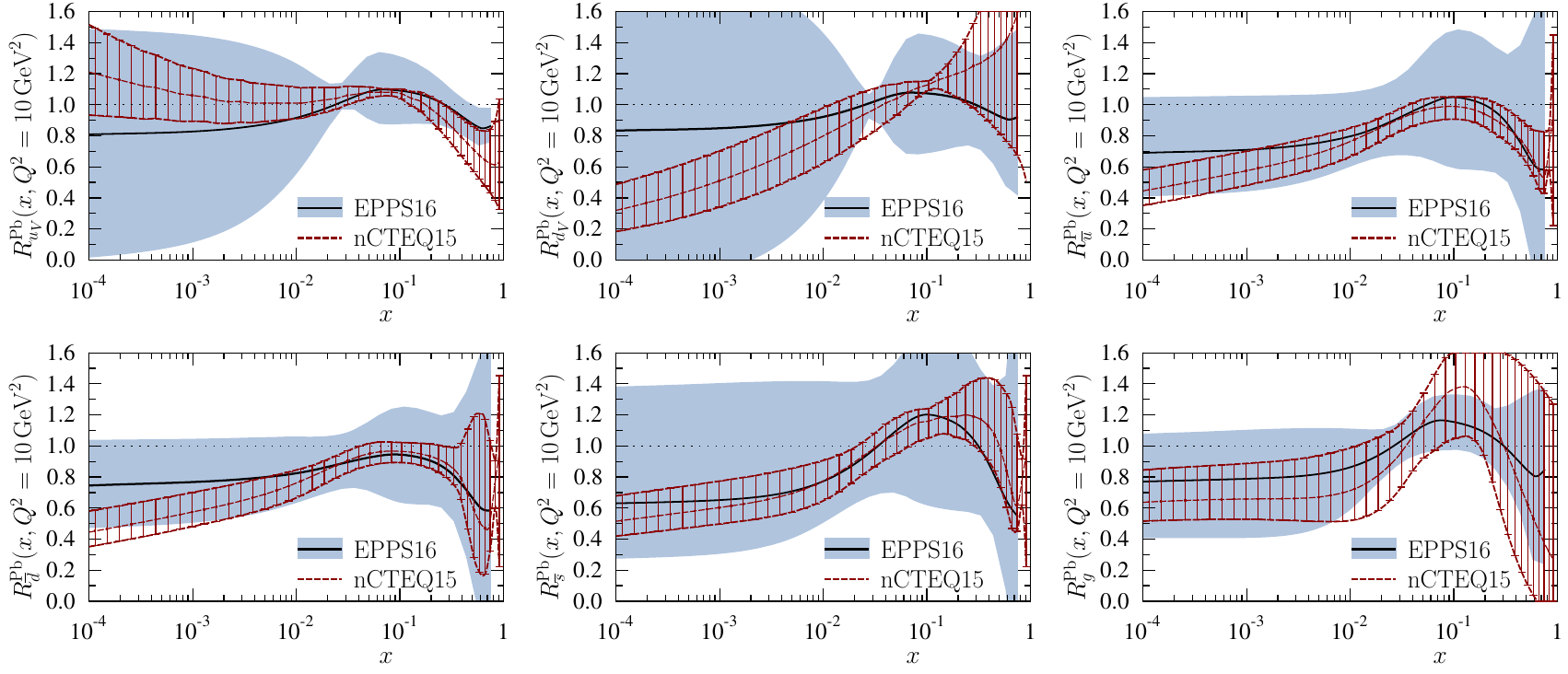}
	\caption{Comparison of nuclear PDF modifications at $Q^2=10$ GeV$^2$ as parametrised in nCTEQ15 (red curves with hatching) \cite{Kovarik:2015cma} and EPPS16 (black central curve with shaded uncertainty bands) \cite{Eskola:2016oht}. Figure taken from Ref.\ \cite{Eskola:2016oht}.}
	\label{fig:npdfs}
\end{figure}

In Fig.\ \ref{fig:npdfs} we show for later reference a comparison of the nuclear PDF modifications
at $Q^2=10$ GeV$^2$ for up/down valence and sea (including strange) quarks as well as gluons as parametrised in nCTEQ15 (red curves with hatching) \cite{Kovarik:2015cma} and EPPS16 (black central curve with shaded uncertainty bands) \cite{Eskola:2016oht}. Note that nPDFs are continuously being updated to include in particular more LHC \cite{Brandt:2014vva,Kusina:2017gkz,Guzey:2019kik,Eskola:2019bgf,Kusina:2020lyz} and JLab data \cite{Segarra:2020gtj}
and NNLO corrections \cite{AbdulKhalek:2019mzd,Walt:2019slu}.

\subsubsection{Hard scattering and scale uncertainties}

At LO, large uncertainties in the hard scattering cross section are introduced by the renormalisation and factorisation procedures. We estimate their numerical size with the traditional seven-point method, i.e.\ by varying the renormalisation scale $\mu_{\rm R}$ and the factorisation scale $\mu_{\rm F}$ independently by relative factors of two, but not four about the central scale, set by the transverse momentum of the underlying hard $2\to2$ event.

\begin{figure} %[htb]
	\centering
	\includegraphics[width=\textwidth]{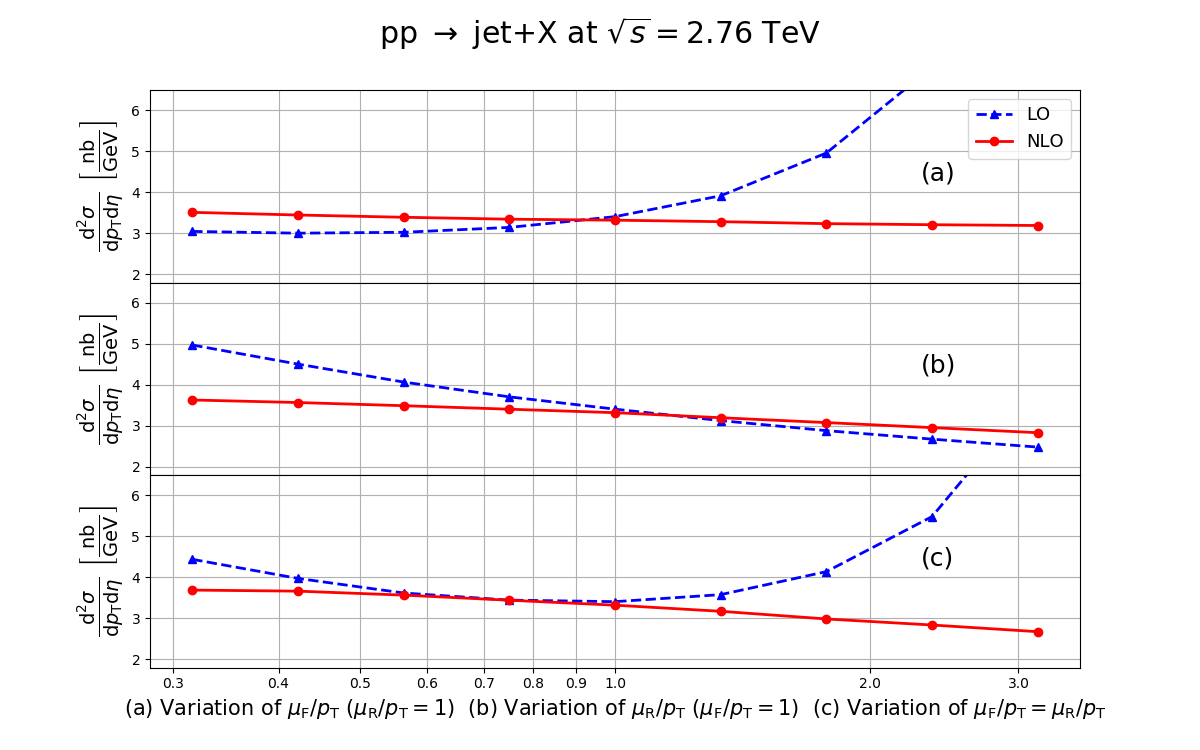}
	\caption{Dependence of the differential cross section for inclusive jets with radius $R=0.4$, transverse momentum $p_{\rm T}\in[70;80]$ GeV and rapidity $|\eta|<2.0$ on the factorisation scale (top), renormalisation scale (centre) and both scales (bottom) at LO (blue curves) and NLO (red curves).
%mkl Events with $n$~jets satisfying the acceptance cuts are counted $n$~times.
%mkl++
          All jets within one event that satisfy the acceptance cuts contribute to the
          cross section.}
%mkl--        
	\label{fig:nloscale}
\end{figure}

The impact of individual variations of $\mu_{\rm F}$ (top panel) and $\mu_{\rm R}$ (central panel) at LO (blue curves) on the differential jet cross section in pp collisions at the LHC with a centre-of-mass energy of $\sqrt{s}=2.76$ TeV is shown in Fig.\ \ref{fig:nloscale}, together with their combined effect (bottom panel). The jet definition, radius and acceptance cuts used are related to those of the CMS measurement (see below). 
At LO, the cross section increases in the range $\mu_{\rm F}/p_{\rm T}\in[0.5;2]$ by about a factor of two, due to the positive scaling violation of the gluon PDFs, which dominate at the low $p_{\rm T}$ values of 70 to 80 GeV (and correspondingly low values of Bjorken $x$) considered here. Conversely, the LO cross section decreases in the range $\mu_{\rm R}/p_{\rm T}\in[0.5;2]$ by almost a factor of two, due to the running of the strong coupling constant. The bottom panel demonstrates that due to this opposite dependence a simultaneous variation of both $\mu_{\rm F}/p_{\rm T}=\mu_{\rm R}/p_{\rm T}\in[0.5;2]$ leads only to small changes in the LO cross section and would thus severely underestimate its error. The total uncertainty would, however, be overestimated with relative factors of four between $\mu_{\rm F}/p_{\rm T}=0.5$ and $\mu_{\rm R}/p_{\rm T}=2$ on the one hand and $\mu_{\rm F}/p_{\rm T}=2$ and $\mu_{\rm R}/p_{\rm T}=0.5$ on the other hand.

For jet production in pp collisions, also NLO predictions matched to parton showers and including hadronisation can be made, e.g.\ with POWHEG \cite{Alioli:2010xa}. The NLO corrections to the hard scattering processes are known to significantly reduce the unphysical scale uncertainties \cite{Klasen:1994bj,Klasen:1996yk}. This is demonstrated as well in Fig.\ \ref{fig:nloscale} (red curves). The differential cross section now decreases monotonically with $\mu_{\rm F}$ and $\mu_{\rm R}$ and by only 10\% and 20\%, respectively, and thus also, when both scales are varied together. The central scale $p_{\rm T}$ is an optimal choice, as LO and NLO predictions agree there. We will come back to this point below when we discuss the scale uncertainties of various jet transverse momentum distributions.

\subsubsection{Medium geometry and parameters}

In JEWEL the Glauber model is used to relate centrality and impact parameter $b$. It is also employed to compute the density of binary nucleon-nucleon collisions $N_\mathrm{coll}$ and number of participants $N_\mathrm{part}$ in the transverse $(x,y)$ plane. The initial parton scatterings take place at $t = z = 0$ and are distributed in the transverse plane according to $N_\mathrm{coll}$. 

Furthermore, the jet yields generated in hard scattering events by JEWEL need to be converted to properly normalised yields or cross sections that can be compared directly to the experimental results in pp or PbPb collisions. Depending on the minimum momentum transfer required in the hard scattering, a certain cross section $\sigma_\mathrm{hard}$ is sampled, representing a fraction of the total inelastic nucleon-nucleon cross section $\sigma_\mathrm{inel}^\mathrm{NN}$. Accounting for the increased number of scatterings in the overlap of nuclei, one obtains the correct normalisation of the JEWEL events to the total number of PbPb collisions in a given centrality
\begin{equation}
    N_\mathrm{Evt} =   \frac{N_\mathrm{hard}}{\langle T_\mathrm{AA} \rangle \cdot \sigma_\mathrm{hard}} =  
    \frac{N_\mathrm{hard}}{ \langle N_\mathrm{coll} \rangle  } \frac{\sigma_\mathrm{inel}^\mathrm{NN}}{ \sigma_\mathrm{hard}},
\end{equation}
where $N_\mathrm{hard}$ is the number of generated events and $N_\mathrm{coll} = T_\mathrm{AA} \cdot \sigma_\mathrm{inel}^\mathrm{NN}$.  
Following Ref.~\cite{Loizides:2017ack} we employ $\sigma_\mathrm{inel}^\mathrm{NN}=61.8$ mb for $\sqrt{s_\mathrm{NN}}=2.76$ TeV and $\sigma_\mathrm{inel}^\mathrm{NN}=67.6$ mb for $\sqrt{s_\mathrm{NN}}=5.02$~TeV. We neglect the normalisation uncertainties arising from the uncertainties of $\sigma_\mathrm{inel}^\mathrm{NN}$ of 0.9 mb and 0.6 mb, respectively.

\subsection{Medium evolution}

The medium evolution in JEWEL is determined by two parameters, the initial temperature $T_{\rm I}$ and the formation time $\tau_{\rm I}$, when the hydrodynamic evolution starts. The initial transverse temperature profile is fixed from the distribution of the participant nucleons in the transverse plane, assuming that the initial energy density $\epsilon \propto T_{\rm I}^4$ is proportional to the density of participants. Starting from $\tau_{\rm I}$ a 1D Bjorken expansion and an ideal-gas equation of state is assumed. For a given centrality or impact parameter $b$, the temperature in the transverse direction changes as
\begin{equation}
 T(x,y,\tau) = T_{\rm I}(x,y,\tau_{\rm I}) \cdot \left(\frac{\tau}{\tau_{\rm I}}\right)^{-1/3}. 
\end{equation}
The transverse expansion, which would lead to faster cooling of the medium, is neglected. In JEWEL, $T_{\rm I}$ therefore reflects the average value of the initial temperatures across the transverse plane. The value of the central cell, which is usually quoted in hydrodynamic models, is thus larger, for instance for central PbPb collisions at $\sqrt{s_\mathrm{NN}}= 2.76$ TeV by about a factor of 1.35. The default values in JEWEL of $T_{\rm I} = 360$ MeV = 485/1.35 MeV and $\tau_{\rm I} = 0.6$\,fm are chosen to match the values in Ref.\ \cite{Shen:2012vn}.

Before the hydrodynamic evolution, at earlier times than $\tau_{\rm I}$, a linear increase of the temperature is implemented in JEWEL
\begin{equation}
T(x,y,\tau < \tau_{\rm I}) = T_{\rm I}(x,y,\tau_{\rm I}) \cdot \left(\frac{\tau}{\tau_{\rm I}}\right) 
\end{equation}
which effectively reduces the sensitivity to the exact choice of $\tau_{\rm I}$.
%mkl++
In Ref.\ \cite{Zapp:2012ak}, a variation of the formation time was found to lead to a $\sim$20\%
variation of the inclusive jet rate in central PbPb collisions.
%mkl--

The hydrodynamic evolution ceases at the crossover temperature $T_{\rm C}$, which has a default value of 170 MeV in JEWEL, consistent with the 1D Bjorken expansion. No hadronic phase is considered.
We will investigate the sensitivity of the model results on both $T_{\rm I}$ and $T_{\rm C}$.

\subsection{Jet evolution}

To the initial hard-scattered partons, the medium appears as a collection of free partons with a statistical momentum distribution according to the current temperature of the medium. The re-scatterings with the medium partons are calculated within JEWEL with standard perturbative techniques, employing LO matrix elements with an additional thermal scale $\mu_{\rm D} = 0.9 \cdot 3 T$ to regulate infrared divergences.
%mkl++
The effect of varying this parameter has been studied for single–inclusive hadron production
in Ref.\ \cite{Zapp:2012ak}.
%mkl--
After the shower has evolved through the medium and vacuum, the event is handed back to PYTHIA 6.4 for hadronisation and decays of unstable particles. Subsequent hadronic rescatterings are not considered (see, however, Ref.\ \cite{Dorau:2019ozd} for a recent study). Jets are reconstructed with the anti-$k_\mathrm{T}$ \cite{Cacciari:2008gp} algorithm as implemented in FastJet \cite{Cacciari:2011ma}.

In JEWEL, the medium response to the hard scattering can be included by keeping the recoiling medium partons in the event record. Since the soft background event is not simulated, JEWEL does not provide a full description of the heavy-ion environment and one cannot follow the exact same prescriptions used in real data analysis, in particular concerning the background removal. The impact of the correlated and uncorrelated background depends on the jet observable and analysis technique. In a simulation without recoils, partons scattered out of the medium are removed from the event and no background is present. This also removes the medium-response to the jet, which is important for a detailed study of jet substructure. We therefore include recoils and subtract the known four-momenta of the thermal components from the reconstructed jet following the procedure described in Ref.\ \cite{KunnawalkamElayavalli:2017hxo}. This then allows for realistic comparisons to experimental data including background effects.

\section{Comparison to experimental results}
\label{sec:3}

We now turn to our numerical results and compare our predictions made with JEWEL and PYTHIA 6.4 to experimental data from the LHC.

\subsection{Data selection}

\begin{figure} %[htb]
	\centering
	\includegraphics[width=0.9\textwidth]{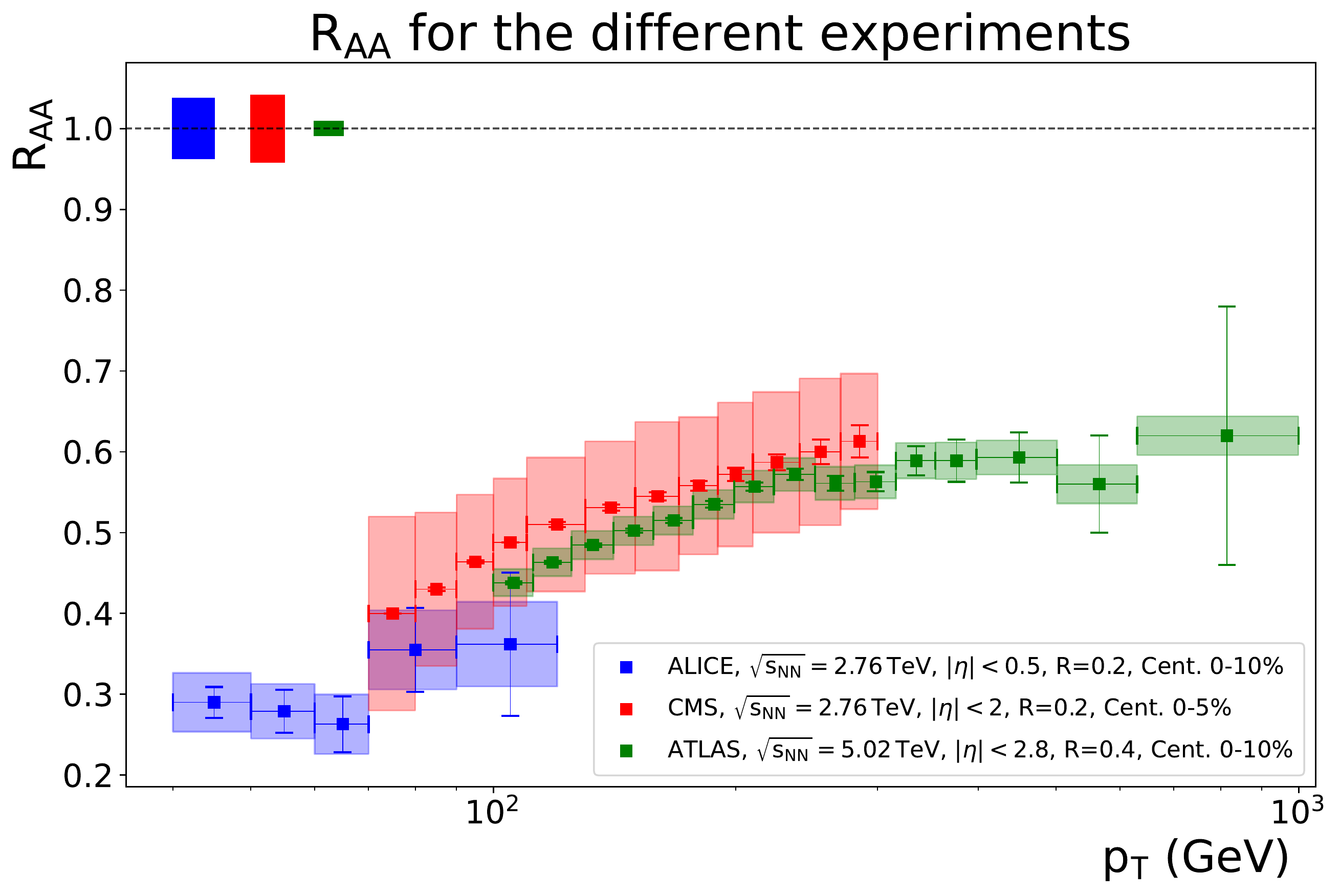}
	\caption{Experimental data sets on jet $R_{AA}$ from ALICE \cite{Abelev:2013fn,Adam:2015ewa}, CMS \cite{Khachatryan:2016jfl} and ATLAS \cite{Aaboud:2018twu} considered in this study. Statistical errors are shown by the error bars, systematic errors are indicated by the shaded boxes, and the overall normalisation uncertainty on $T_{\rm AA}$ is indicated by the boxes around unity.}
	\label{fig:raa_exp}
\end{figure}

The data sets we compare to have been obtained with the anti-$k_{\rm T}$ jet algorithm \cite{Cacciari:2008gp} by the
\begin{itemize}
\item ALICE collaboration in pp \cite{Abelev:2013fn} and PbPb \cite{Adam:2015ewa} collisions at a centre-of-mass energy per nucleon of $\sqrt{s_{\rm NN}}=2.76$ TeV for inclusive jets with radius $R=0.2$, rapidity range $|\eta|<0.5$ and transverse momentum $p_{\rm T}\in[40;120]$ GeV (0-10\% centrality) and $p_{\rm T}\in[30;100]$ GeV (10-30\% centrality)\footnote{In addition, each jet was required to contain a charged particle with $p_\mathrm{T}>5$ GeV, which holds for more than 90\% (95\%) of the jets with $p_{\rm T}>40$ GeV (50 GeV) \cite{Adam:2015ewa}.};
\item CMS collaboration in pp and PbPb collisions \cite{Khachatryan:2016jfl} at a centre-of-mass energy per nucleon of $\sqrt{s_{\rm NN}}=2.76$ TeV for inclusive jets with radii $R=0.2$, 0.3 and 0.4, rapidity range $|\eta|<2$ and transverse momentum $p_{\rm T}\in[70;300]$ GeV (0-5\% centrality);
\item ATLAS collaboration in pp and PbPb collisions \cite{Aaboud:2018twu} at a centre-of-mass energy per nucleon of $\sqrt{s_{\rm NN}}=5.02$ TeV for inclusive jets with radius $R=0.4$, rapidity range $|\eta|<2.8$ and transverse momentum $p_{\rm T}\in[100;1000]$ GeV (0-10\% centrality).
\end{itemize}
The experimental data and their statistical and systematic uncertainties have been retrieved from the HEPData data base \cite{Maguire:2017ypu}, and the analysis cuts and selections specific to each experiment have been implemented on an event-by-event basis using RIVET \cite{Buckley:2010ar}. The experimental results on the jet nuclear modification factor $R_{\rm AA}$ are collected in Fig.~\ref{fig:raa_exp}.
Statistical errors are shown by the error bars, systematic errors are indicated by the shaded boxes, and the overall normalisation uncertainty on $T_{\rm AA}$ is indicated by the boxes around unity. Qualitatively, the measurements are in good agreement despite slightly different jet radii and central rapidity ranges. The largest suppression (quenching) is observed by the ALICE experiment at low $p_{\rm T}\simeq 40$ GeV, where $R_{\rm AA}$ lies around 0.3. The jet nuclear modification factor increases to 0.6 for large $p_{\rm T}\simeq 1$ TeV as measured by the ATLAS experiment.

\subsection{Jet spectra and nuclear modification factors}

\begin{figure} %[htb]
	\centering
	\includegraphics[width=0.8\textwidth]{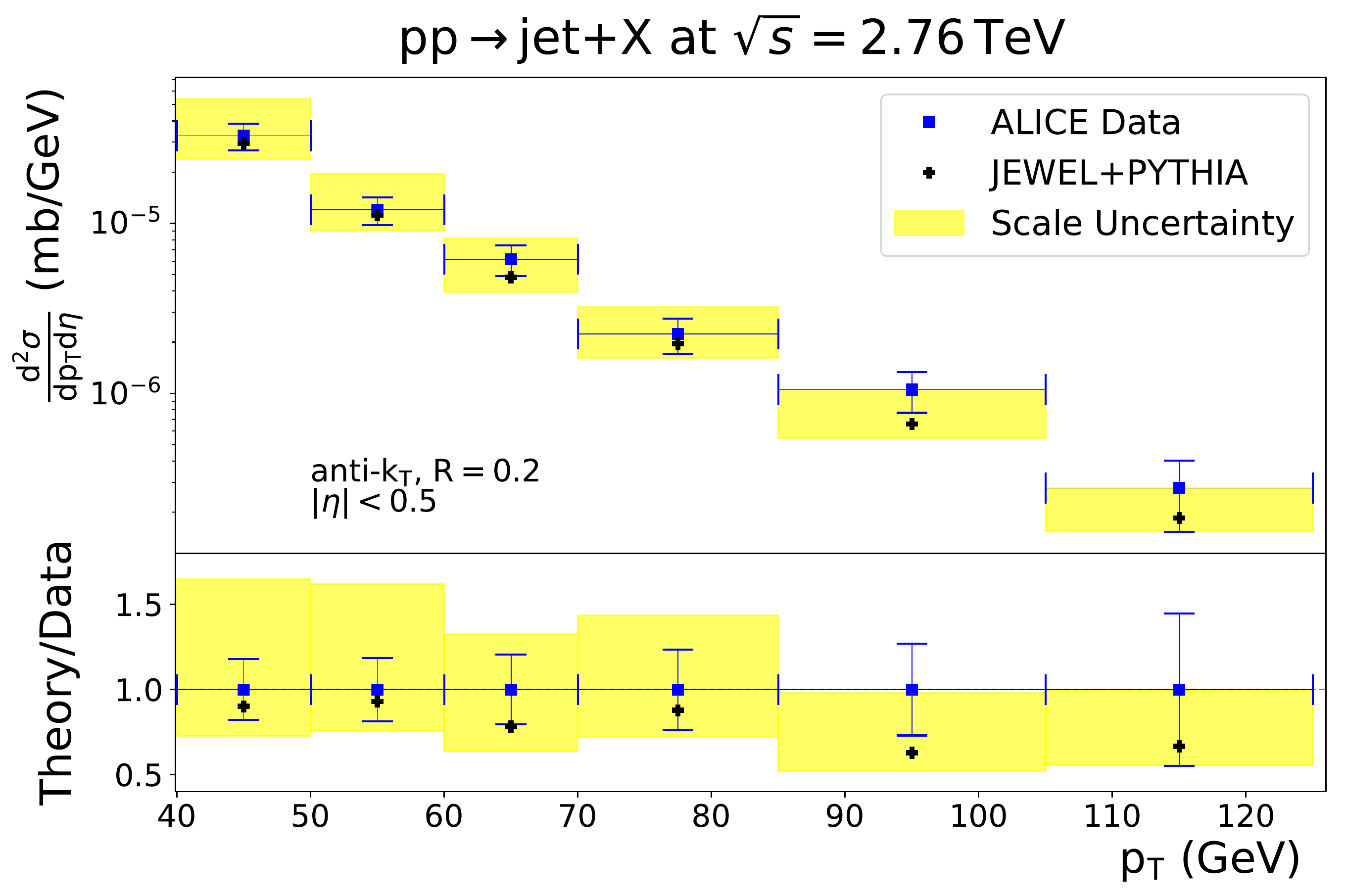}
	\caption{Transverse momentum distribution of jets with radius $R=0.2$ in pp collisions at $\sqrt{s}=2.76$ TeV measured by ALICE in the rapidity range $|\eta|<0.5$ \cite{Abelev:2013fn} and calculated with JEWEL. Scale uncertainties are shown as yellow shaded bands.}
	\label{fig:alicepp}
\end{figure}

The transverse momentum spectrum of inclusive jets with radius $R=0.2$ produced in the rapidity range $|\eta|<0.5$ in pp collisions at $\sqrt{s}=2.76$ TeV is shown in Fig.\ \ref{fig:alicepp}. Data from the ALICE experiment %
\cite{Abelev:2013fn}, where statistical and systematic errors have been added in quadrature, are compared to our predictions with JEWEL, for which the scale uncertainties obtained with the seven-point method described above are shown as yellow shaded bands. The upper panel shows good qualitative agreement which is confirmed quantitatively in the ratio of theory/data in the lower panel. For most of the data points, even the central theoretical values, which NLO calculations with POWHEG prove to be more reliable than the LO scale uncertainties suggest (see above), lie within the total experimental error, which increases due to the growing statistical uncertainty as the event rates fall at large~$p_{\rm T}$.

\begin{figure} %[htb]
	\centering
	\includegraphics[width=\textwidth]{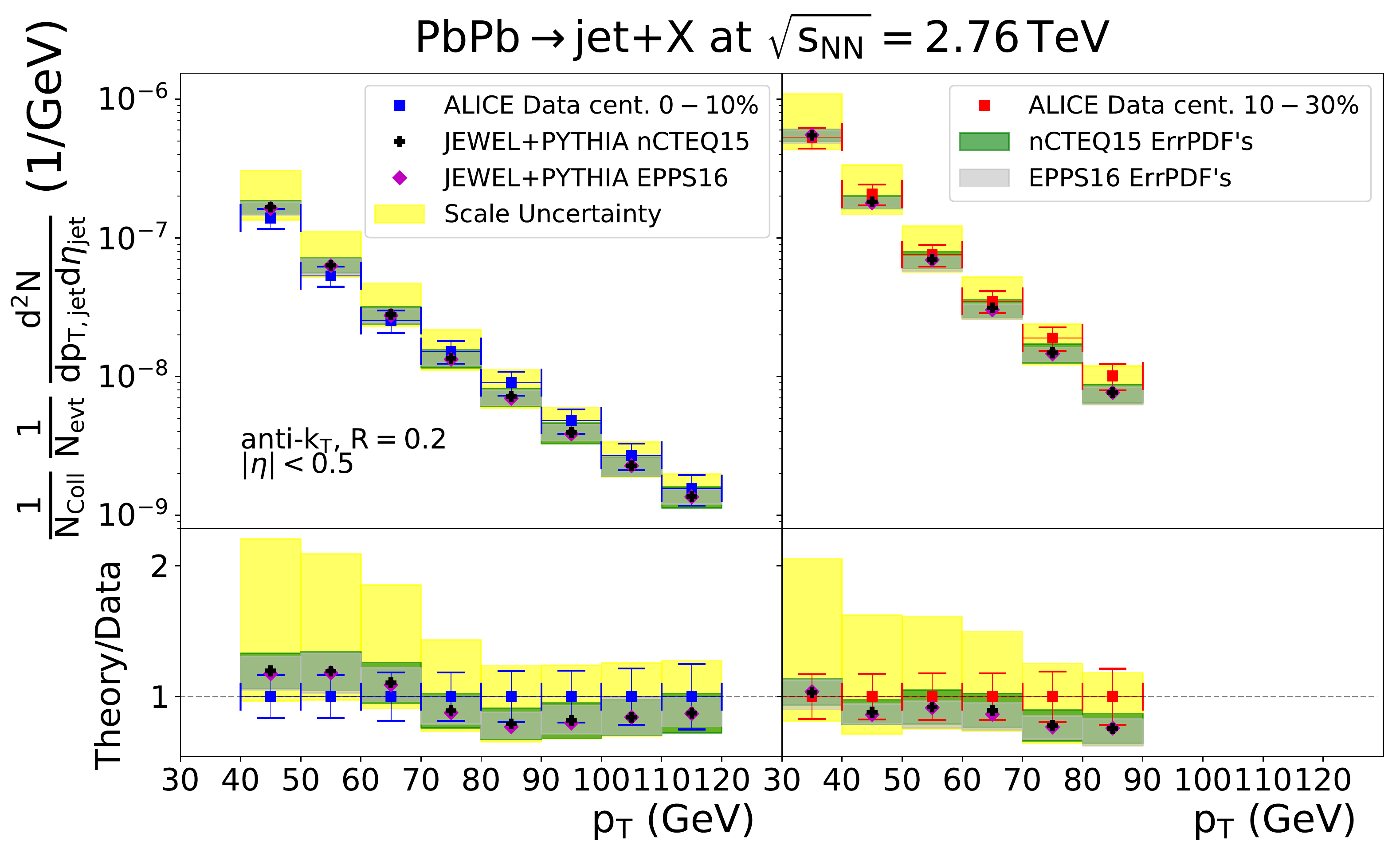}
	\caption{Transverse momentum distributions of jets with radius $R=0.2$ in PbPb collisions at $\sqrt{s_{\rm NN}}=2.76$ TeV measured by ALICE in the rapidity range $|\eta|<0.5$ and the centrality ranges $0-10\%$ (left) and $10-30\%$ (right) \cite{Adam:2015ewa} and calculated with JEWEL. Scale (nuclear PDF) uncertainties are shown as yellow (green and grey) shaded bands.}
	\label{fig:alicepbpb}
\end{figure}

The corresponding measurements in 0-10\% (left) and 10-30\% (right) central PbPb collisions are shown in Fig.\ \ref{fig:alicepbpb}.
In addition to the theoretical scale uncertainties, we show also those induced by variations of the nuclear PDFs. Qualitatively we find good agreement in the upper panels, which is again confirmed quantitatively in the ratio theory/data in the lower panels. As before, the central theoretical values lie within the experimental errors for most of the points. This demonstrates that the interaction of the jet with the medium is well described within JEWEL. The nuclear PDF uncertainties are fairly constant and amount to about $\pm15$\% as estimated both by nCTEQ15 (green) and EPPS16 (grey shaded bands). An estimate of the relevant initial longitudinal parton momentum fraction with $x_{\rm T}=2p_{\rm T}/\sqrt{s_{\rm NN}}\in[0.02;0.09]$ and a look at Fig.\ \ref{fig:npdfs} confirm that we are in the transition region from shadowing to anti-shadowing, where these uncertainties are relatively small. For pp, 0-10\% and 10-30\% central PbPb collisions, we observe a tendency for the theory/data ratio to decrease slightly from small to large transverse momenta.

\begin{figure} %[htb]
	\centering
	\includegraphics[width=\textwidth]{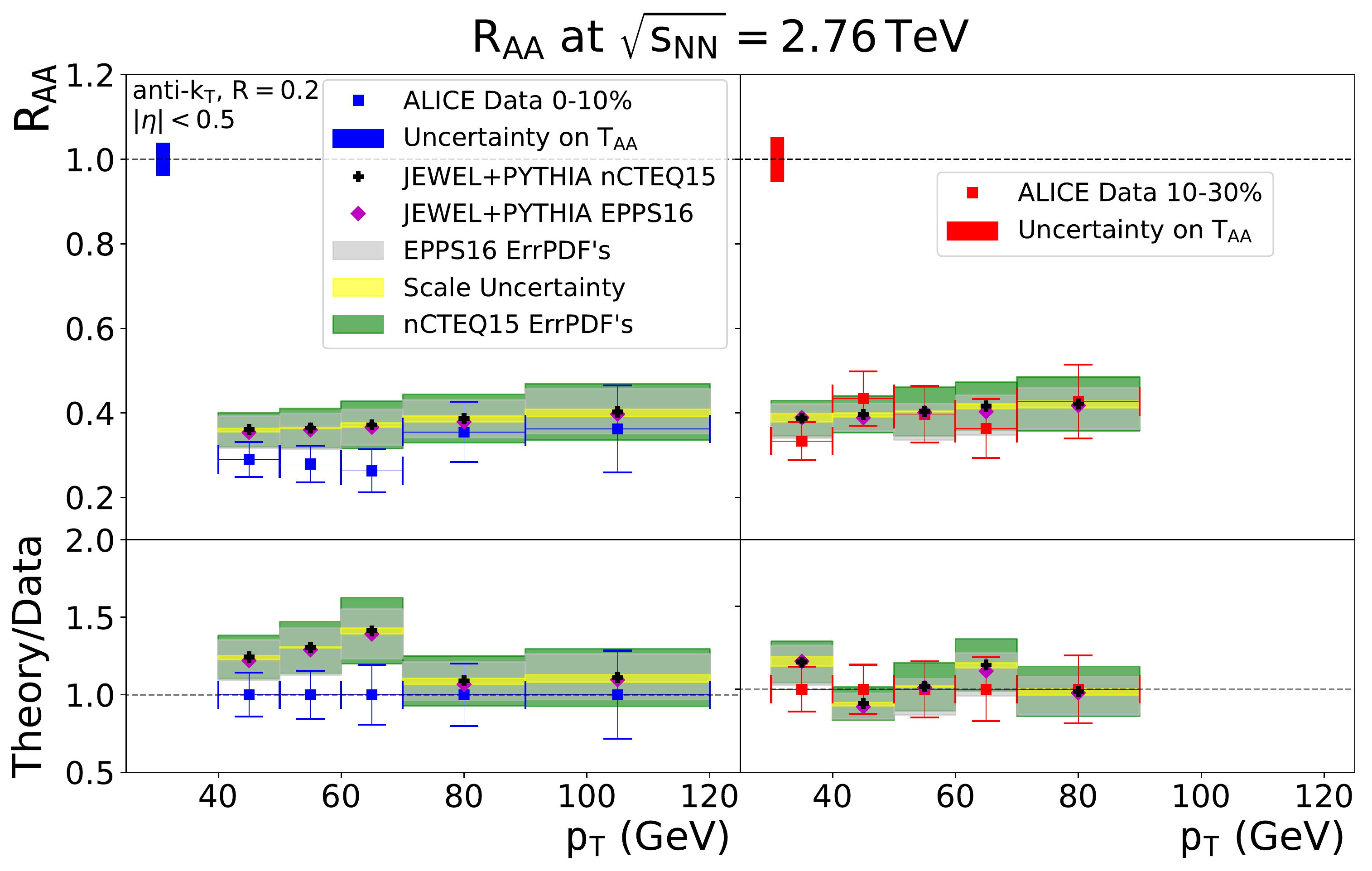}
	\caption{Nuclear modification factor $R_\mathrm{AA}$ for jets with radius $R=0.2$ at $\sqrt{s_{\rm NN}}=2.76$ TeV measured by ALICE in the rapidity range $|\eta|<0.5$ and the centrality ranges $0-10\%$ (left) and $10-30\%$ (right) \cite{Adam:2015ewa} and calculated with JEWEL. Scale (nuclear PDF) uncertainties are shown as yellow (green and grey) shaded bands. The overall normalisation uncertainty on $T_{\rm AA}$ is indicated by the boxes around unity.}
	\label{fig:aliceraa}
\end{figure}

Dividing the jet yields in PbPb collisions by the corresponding cross sections in pp collisions as in Eq.\ (\ref{eq:RAA}) leads to the classic observable for jet quenching, the nuclear modification factor $R_{\rm AA}$, which we show for the ALICE experiment and 0-10\% (left) as well as 10-30\% (right) central collisions in Fig.\ \ref{fig:aliceraa}.
As expected from our previous discussions, the upper panels show good qualitative agreement of theory and data, i.e.\ both the amount of jet quenching of about 0.3 at low $p_{\rm T}$ and the rise to 0.4 at larger $p_{\rm T}$ observed by ALICE are well reproduced by JEWEL. Quantitatively, in the (double) ratios of theory/data (lower panels) the central theoretical predictions agree well with the data in the case of 10-30\% central collisions, but only for the higher $p_{\rm T}$ bins in the case of 0-10\% central collisions. Good agreement is, however, also found for the lower $p_{\rm T}$ bins once the nuclear PDF uncertainties are accounted for, which underlines their importance. 
The scale uncertainty in the ratio is evaluated again as the spread of the ratios with the same seven-point-variations as before. It can be seen that the uncertainty cancels to a large extent and is much smaller than the PDF uncertainty (cf.\ also Ref.\ \cite{Huss:2020dwe}). 

\begin{figure}[htb]
	\centering
	\includegraphics[width=\textwidth]{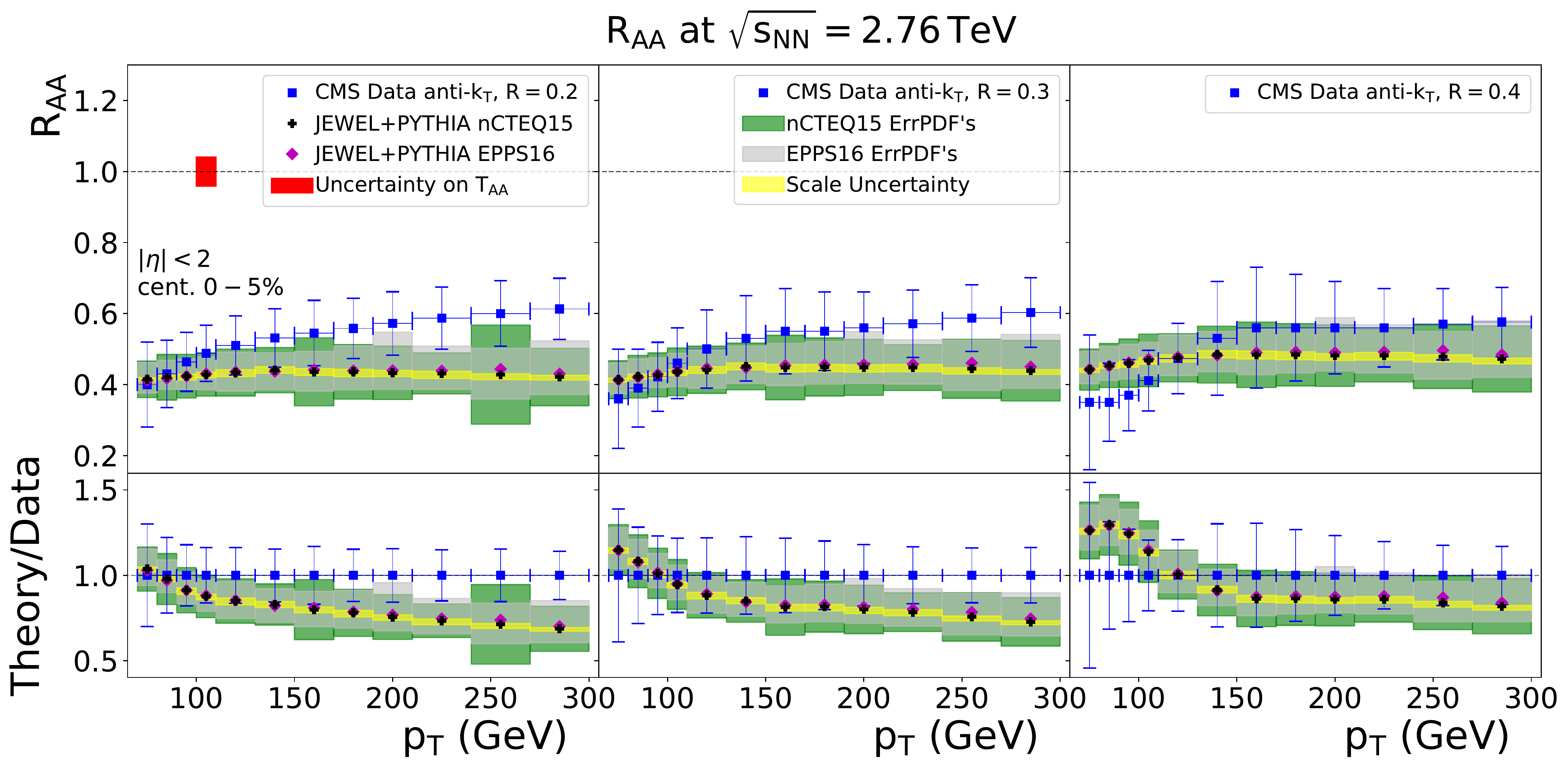}
	\caption{Nuclear modification factor $R_\mathrm{AA}$ for jets with radii $R=0.2$ (left), 0.3 (centre) and 0.4 (right) at $\sqrt{s_{\rm NN}}=2.76$ TeV measured by CMS in the rapidity range $|\eta|<2$ and the centrality range $0-5\%$ \cite{Khachatryan:2016jfl} and calculated with JEWEL. Scale (nuclear PDF) uncertainties are shown as yellow (green and grey) shaded bands. The overall normalisation uncertainty on $T_{\rm AA}$ is indicated by the box around unity.}
	\label{fig:cmsraa}
\end{figure}

With the CMS experiment, also jets with larger transverse momenta (up to 300 GeV) and in a larger rapidity range ($|\eta|<2$) than those measured in the ALICE experiment (up to 120 GeV and $|\eta|<0.5$) can be observed. In addition, the CMS collaboration has published results for three different jet radii, i.e.\ $R=0.2$, 0.3 and 0.4. While the centre-of-mass energies are the same, the centrality range in PbPb collisions is reduced in the CMS (0-5\%) compared to the most central ALICE (0-10\%) measurement. Despite these minor differences, the nuclear modification factors measured by ALICE and CMS were found to be in qualitative agreement in the $p_{\rm T}$ overlap region in Fig.\ \ref{fig:raa_exp}. Fig.\ \ref{fig:cmsraa} (left, upper panel) shows 
the CMS data for $R=0.2$ again, but now in comparison with our JEWEL predictions. While the description of the data is again very good for low $p_{\rm T}\leq150$ GeV, the rise of the data at large $p_{\rm T}\simeq 300$ GeV is not entirely reproduced by JEWEL. This is also clear from the (double) ratio of theory/data in the lower panel, where the central values drop below one as the jet transverse momentum increases. Taking into account the nuclear PDF uncertainties (green and grey shaded bands), we can, however, account for the data within errors up to the largest $p_{\rm T}$. The rise of $R_{\rm AA}$ in the data reduces when the jet radius $R$ is increased to 0.3 (center) and in particular for 0.4 (right), where also the central theoretical predictions describe all data points within the experimental error alone. This implies that not only the initial parton distributions, but also the structure and evolution of the final jet in the medium are still not entirely known and deserve further study.

\begin{figure}
	\centering
	\includegraphics[width=0.85\textwidth]{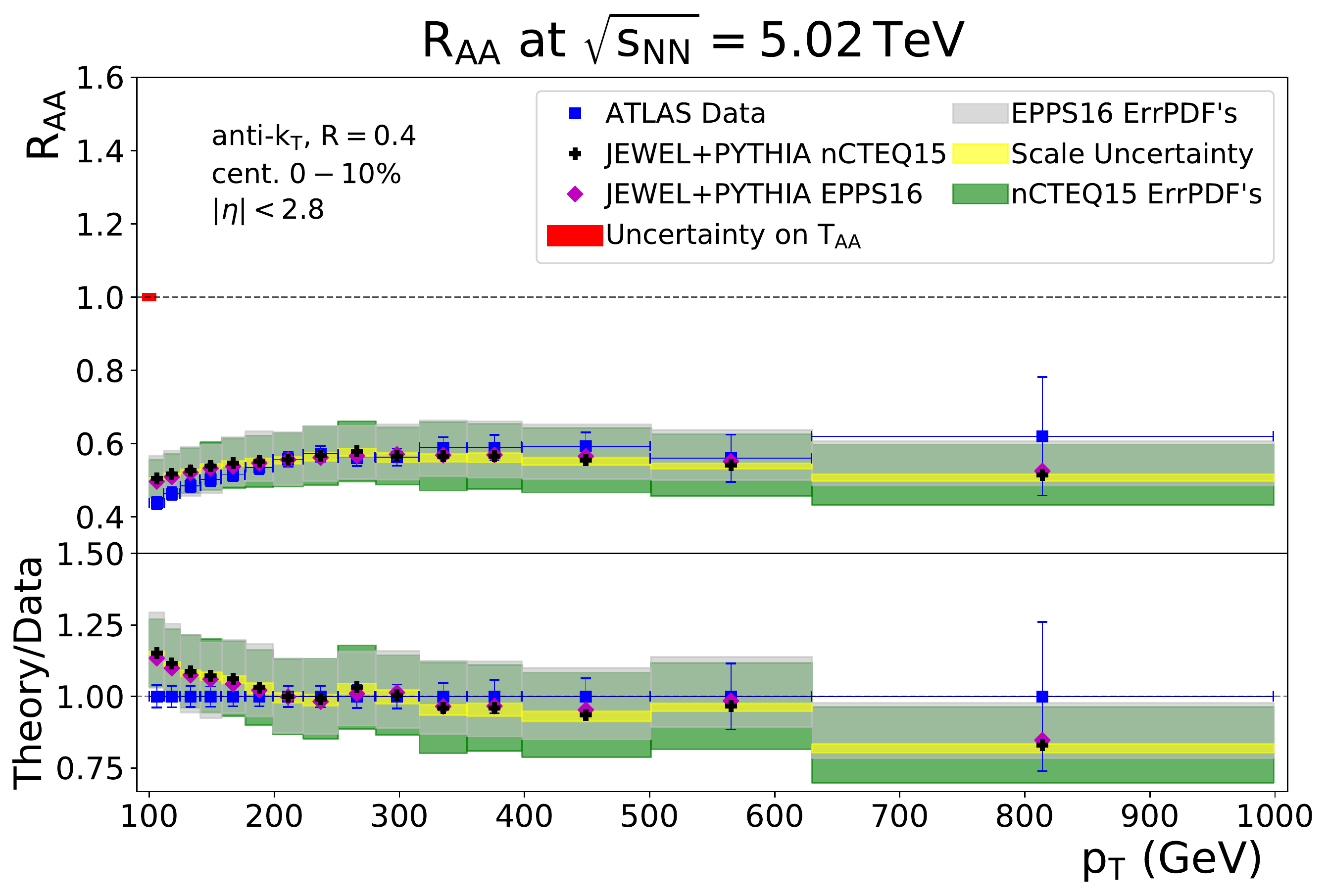}
	\caption{Nuclear modification factor $R_\mathrm{AA}$ for jets with radius $R=0.4$ at $\sqrt{s_{\rm NN}}=5.02$ TeV measured by ATLAS in the rapidity range $|\eta|<2.8$ and the centrality range $0-10\%$ \cite{Aaboud:2018twu} and calculated with JEWEL. Scale (nuclear PDF) uncertainties are shown as yellow (green and grey) shaded bands. The overall normalisation uncertainty on $T_{\rm AA}$ is indicated by the box around unity.}
	\label{fig:atlasraa}
\end{figure}

The ATLAS measurement has been obtained with the same large jet radius $R=0.4$ as the third CMS measurement, but extends to even higher $p_{\rm T}$ of 1000 GeV. At the same time, it has been performed at a higher centre-of-mass energy per nucleon of $\sqrt{s_{\rm NN}}= 5.02$~TeV. From the larger jet radius, we expect again a leveling off of $R_{\rm AA}$, but at only higher~$p_{\rm T}$ due to the larger collision energy. This is indeed observed in the measurement shown in Fig.\ \ref{fig:atlasraa} (upper panel), where $R_{\rm AA}$ rises to 0.6 at $p_{\rm T}=300$ GeV, but then stays constant. The slightly larger rapidity ($|\eta|<2.8$) and centrality 
(0-10\%) ranges should not have a significant impact on this conclusion. The comparison of the JEWEL prediction to the data is best done in the (double) ratio theory/data (lower panel). The trend at low $p_{\rm T}$ of the central theory predictions to lie above the data, that was already present in the $R=0.4$ CMS data, is confirmed, but for most of the data points the agreement is excellent. If we take nuclear PDF uncertainties into account, all data points can be described.

\subsection{Dependence on medium parameters}

While our understanding of the thermal properties of QGP matter and the flow of the nearly perfect fluid has
recently made significant progress \cite{Romatschke:2017ejr,Busza:2018rrf}, the theoretical description of the early stages of the heavy-ion collision is still an outstanding problem. In particular, the rise of the nPDFs at low $x$, in particular the one of the gluon which dominates there, is geometrically expected to be limited by recombination and saturation \cite{Albacete:2014fwa}. First experimental evidence for saturation in HERA \cite{Rezaeian:2012ji,Golec-Biernat:2017lfv} and LHC \cite{Levin:2010dw,Dusling:2012cg} data has been claimed, but is still under debate \cite{Mantysaari:2018nng}. The high gluon density at low $x$ should eventually lead to collective behavior, thermalisation and the creation of the QGP with an initial temperature $T_{\rm I}$ \cite{Berges:2020fwq}. This temperature can in principle be extracted from measurements of prompt photon production, since the photons produced at early stages of the collision should not interact with the medium. However, photons are emitted at all stages of the collision, i.e.\ also during the cooling process and even by charged hadrons after the final phase transition, so that in reality only an effective temperature $T_{\rm eff}\simeq300$ MeV can be extracted \cite{Klasen:2013mga,Adam:2015lda}.

\begin{figure} %[htb]
	\centering
	\includegraphics[width=\textwidth]{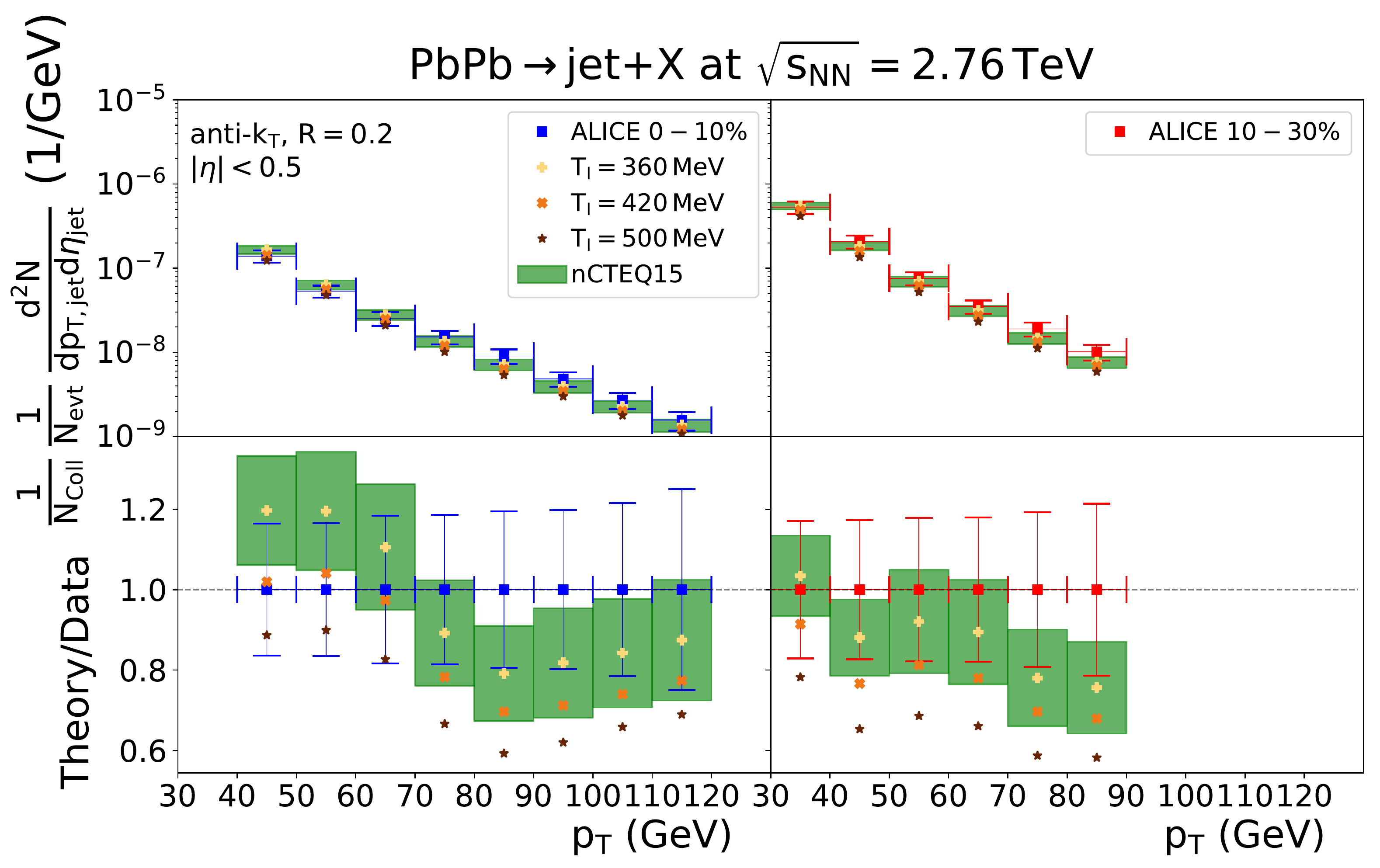}
	\caption{Dependence of the jet yield in 0-10\% (left) and 10-30\% (right) central PbPb collisions as observed by ALICE at $\sqrt{s_{\rm NN}}=2.76$ TeV on the initial temperature $T_\mathrm{I}$ of the QGP. The default value in JEWEL for $T_\mathrm{I}$ is $\SI{360}{MeV}$. The PDF variations correspond to the green shaded band.}
	\label{fig:aliceti}
\end{figure}

A detailed analysis of the initial phase transition from partons confined in heavy nuclei to the QGP is of course beyond the scope of this work. However, it is an interesting question whether correlations between the initial gluon density and the initial temperature can be observed. We therefore show in Fig.\ \ref{fig:aliceti} again (see Fig.\ \ref{fig:alicepbpb}) the jet yield in 0-10\% (left) and 10-30\% (right) central
PbPb collisions as observed by ALICE at $\sqrt{s_{\rm NN}}=2.76$ TeV and compare it to JEWEL predictions assuming three different initial temperatures: the default value 360 MeV and two higher temperatures of 420 and 500 MeV. In Fig.\ \ref{fig:aliceti} (left), we find that for a low initial temperature of $T_{\rm I}=360$ MeV, the data, e.g.\ in the three lowest $p_{\rm T}$ bins, are best described by the lower edge of the nPDF uncertainty band. At the same date, the data for central collisions are also well described by the central nPDF set and a higher initial temperature of 420 MeV. In other words, a good description of the data requires either a low gluon density leading to a smaller hard scattering cross section, or a higher temperature leading to more suppression in the medium. These two effects can only be disentangled, if measurements at different centralities are available. If we therefore now compare the two centralities (left and right) in the three lowest $p_{\rm T}$ bins, we clearly find a lower initial temperature (360 MeV) for the 10-30\% central collisions than for the 0-10\% collisions (420 MeV) as expected. Since a higher temperature implies a longer evolution of the medium and thus more suppression, this corresponds also well to the smaller $R_{\rm AA}\sim0.3$ observed in Fig.\ \ref{fig:aliceraa} for 0-10\% collisions (left) than those of $\sim0.4$ observed for 10-30\% collisions (right). A point to remember, though, is that the JEWEL medium is based on a Bjorken model and thus a one-dimensional expansion, which corresponds to a rather slow cooling of the medium. A full three-dimensional expansion would induce a faster cooling and might reduce the effect of the temperature variation. Looking now at the data at larger $p_{\rm T}$ in both the 0-10\% and 10-30\% collisions, we find that the data require either an increase in gluon density or a decrease in temperature with the transverse momentum. For the 0-10\% data, this means that either both shadowing and anti-shadowing effects - related by the momentum sum rule - are underestimated in the central nCTEQ15 fit, or that the one-dimensional Bjorken model in JEWEL overestimates the interaction of the medium with faster jets. A comparison of the shapes of the transverse-momentum distributions for the two centralities (left and right), as well as the trend of the high $p_\mathrm{T}$ suppression in comparison to the CMS data  in Fig.\ \ref{fig:cmsraa} suggest the latter interpretation.

Since JEWEL only considers interactions in the deconfined phase, rescattering in the medium is only possible as long as the local temperature is higher than the critical temperature $T_{\rm C}$. At this temperature, QCD matter transfers from partonic to hadronic degrees of freedom. By default, the critical temperature in JEWEL is $T_\mathrm{C}=\SI{170}{MeV}$, while more recently from lattice QCD calculations \cite{Borsanyi:2013bia,Bazavov:2018mes} and statistical analyses of hadron abundances at the LHC \cite{Andronic:2018cr} a temperature of $T_{\rm C} \approx 156$ MeV is found. However, these values cannot be compared directly since for a 1D hydrodynamic expansion typically larger values for $T_C$ are needed to reproduce the bulk features of particle production. 

\begin{figure} %[htb]
	\centering
	\includegraphics[width=\textwidth]{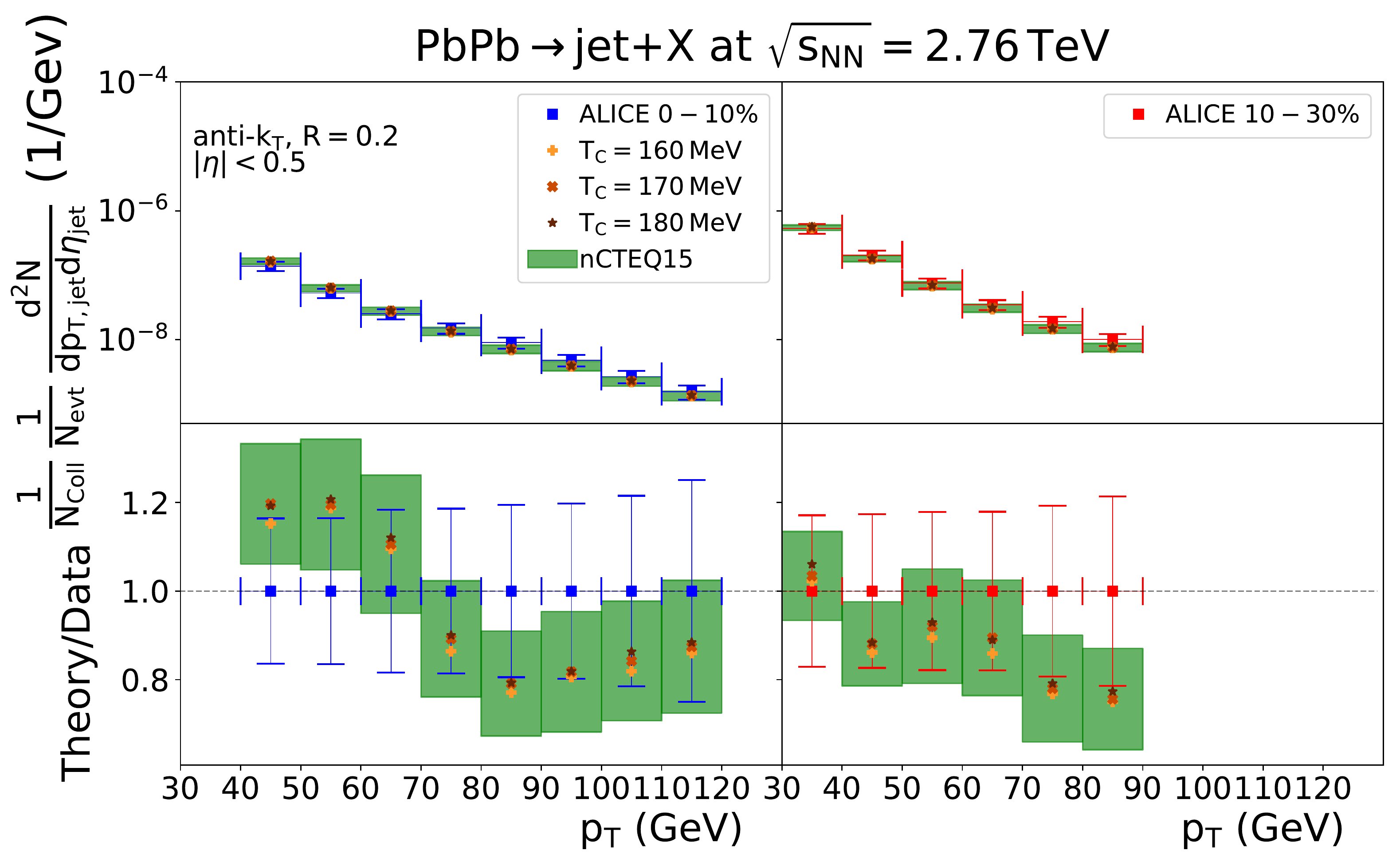}
	\caption{Dependence of the jet yield on the critical temperature $T_\mathrm{C}$ of the QGP. The default value in JEWEL for $T_\mathrm{C}$ is $\SI{170}{MeV}$. The PDF uncertainties correspond to the green shaded band.}
	\label{fig:alicetc}
\end{figure}

We therefore focus on the qualitative effect of variation of $T_C$ and compare in Fig.\ \ref{fig:alicetc} the ALICE data with 
JEWEL calculations using also lower and larger critical temperatures of 160 MeV and 180 MeV. Overall, the dependence on $T_{\rm C}$ is much smaller than the one on~$T_{\rm I}$. We attribute this to the fact that parton-hadron duality is expected to be realised in sufficiently inclusive jet observables \cite{Azimov:1984np} such as the yield of inclusive jets defined by the anti-$k_T$ algorithm \cite{Cacciari:2008gp} and modeled in the Lund string fragmentation model implemented in PYTHIA 6.4 \cite{Sjostrand:2006za}. Lower critical temperatures imply a longer evolution of the jet through the medium and therefore a larger suppression. At low transverse momenta, the results for the (now favoured) value of $T_{\rm C}=160$ MeV indeed agree slightly better with the data for both centralities. In contrast, at medium and large transverse momenta there is almost no dependence, and in any case it is much smaller than the nuclear PDF uncertainty, so that no firm conclusions can be drawn from the inclusive jet yield observable alone.

\subsection{Jet structure and Lund plane}

\begin{figure} %[htb]
    \centering
    \includegraphics[width=.49\textwidth]{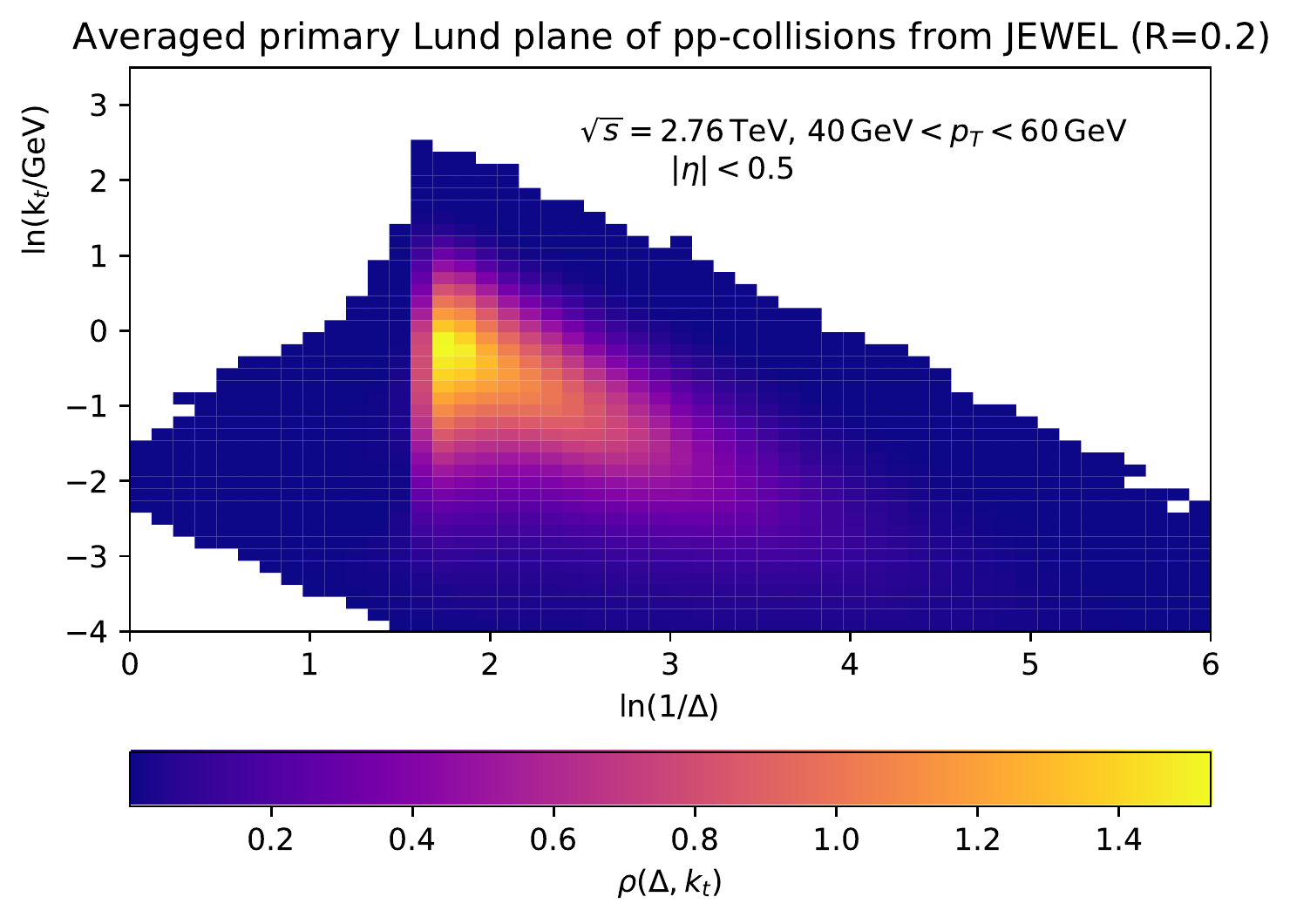}  
    \includegraphics[width=.49\textwidth]{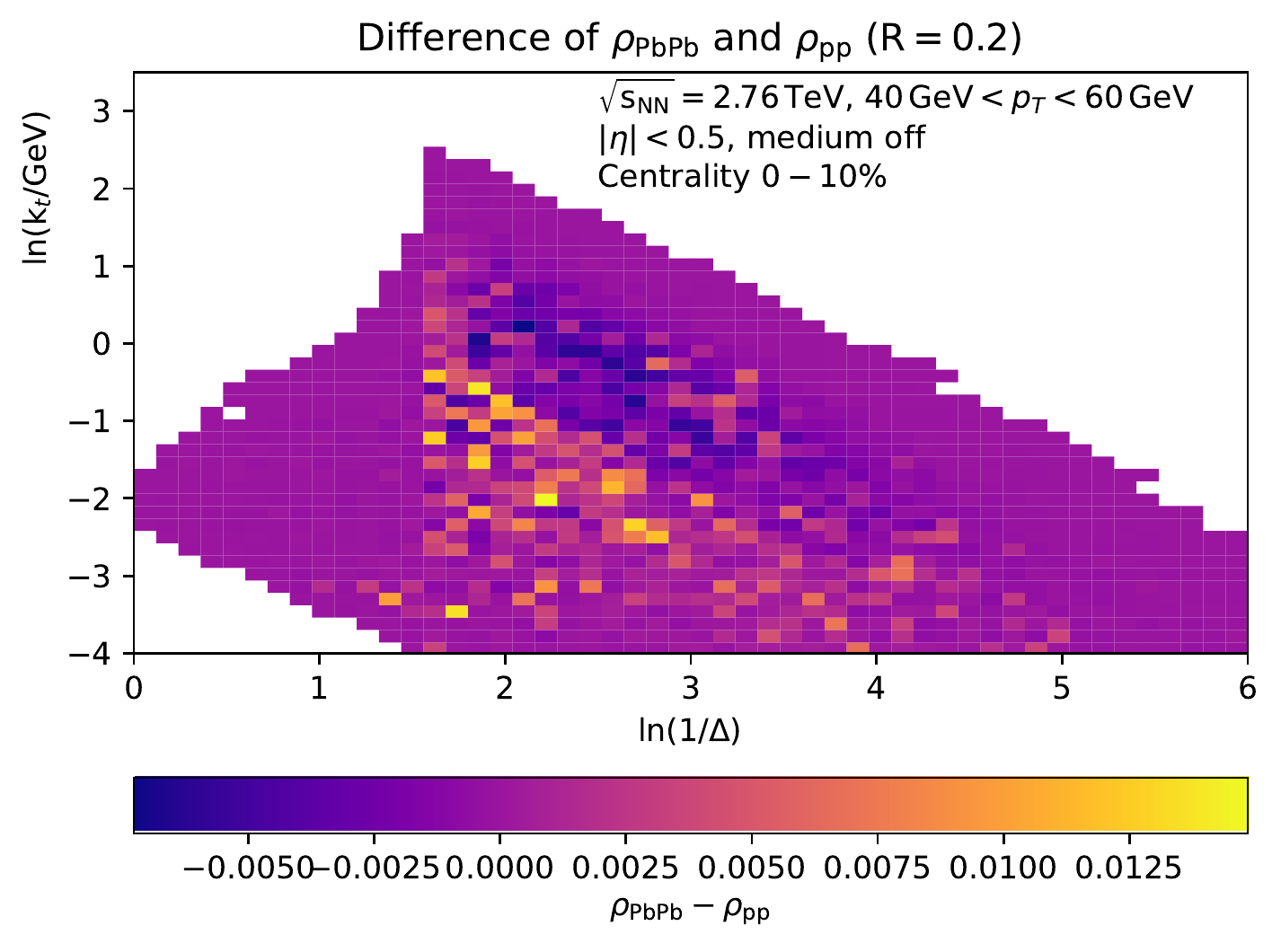}  
    \includegraphics[width=.49\textwidth]{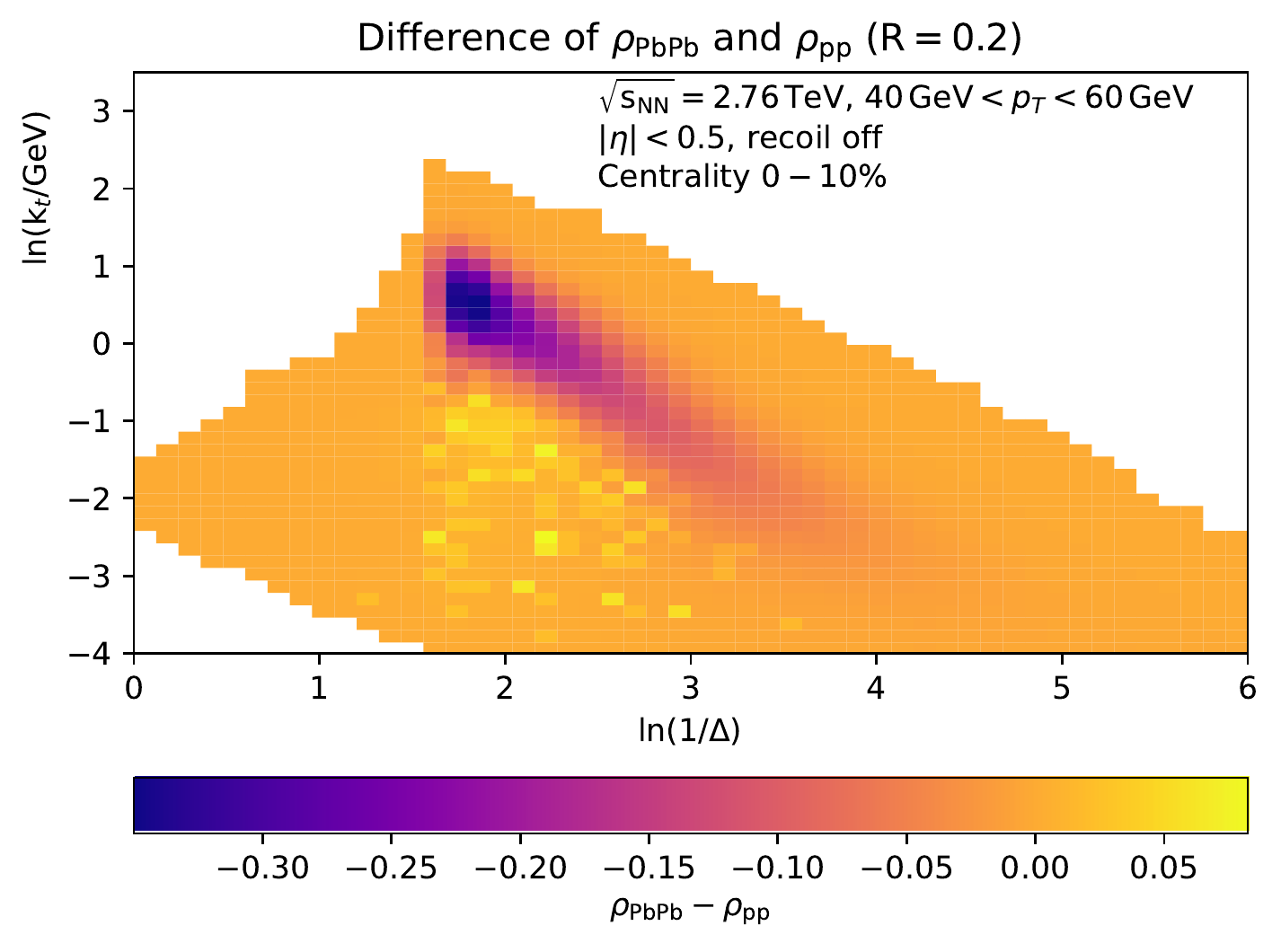}  
    \includegraphics[width=.49\textwidth]{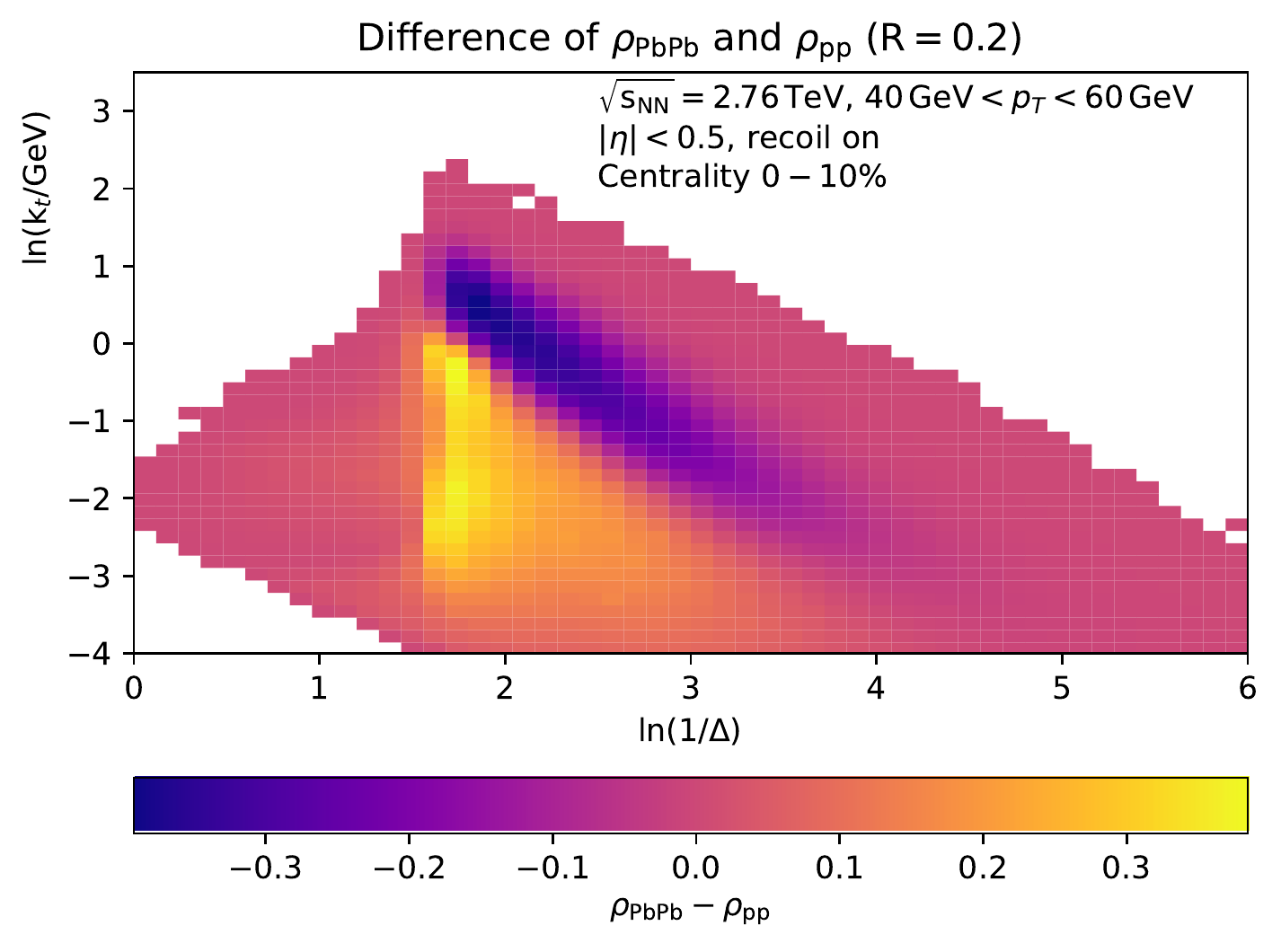}  
    \caption{The average primary Lund plane density $\rho$ for jets simulated with JEWEL, initially clustered with the anti-$k_\mathrm{T}$ algorithm ($R=0.2$,  $40<\hat{p}_\mathrm{T}<\SI{60}{GeV}$ and $|\eta|<0.5$) and then reclustered with the Cambridge/Aachen algorithm. (a) pp collisions, (b) the difference of pp and central PbPb collisions with the medium turned off, (c) the same with the medium turned on and recoil turned off, and (d) the same with both the medium and the recoil turned on.}
    \label{fig:lund_plane}
\end{figure}

To study the evolution of the final-state jet in the medium, it is thus interesting to consider less inclusive quantities like, e.g., jet shapes \cite{Klasen:1997tj,Biswas:2019jkr}. The phase space within jets can be effectively represented in so-called Lund diagrams, where the angle and transverse momentum of an emission with regard to its emitter is mapped onto a triangle in two-dimensional space and each new emission opens up a new phase space, i.e.\ a new triangular leaf, for further emissions \cite{Andersson:1988gp}. While Lund diagrams have been used mainly as a tool in the theoretical discussion of parton showers and resummation \cite{Dasgupta:2013ihk}, they can also be reconstructed experimentally as proposed e.g.\ in Ref.\ \cite{Dreyer:2018nbf}
%mkl++
and for heavy-ion collisions in Ref.\ \cite{Andrews:2018jcm}.
%mkl--

In the following we will use the so-called primary Lund plane of the jets reconstructed within JEWEL events to study the effects of cold and hot nuclear matter on the jet structure.  The simulation is done for $\sqrt{s_{\rm NN}}=\SI{2.76}{TeV}$ and for transverse momenta in the partonic cross section of $\SI{40}{GeV}<\hat{p}_{\rm T}<\SI{60}{GeV}$. The PbPb events are generated  for the 0-10\% most central collisions as discussed above. For a qualitative study of medium-induced effects, three cases will be compared: PbPb events without medium, PbPb events with the recoil option turned off, and PbPb events with the recoil option on. The primary Lund plane, i.e.\ the first triangular plane, is built from jets reconstructed with the anti-k$_\mathrm{T}$ algorithm and radius $R = 0.2$. All jets with $p_{\rm T}>\SI{10}{GeV}$ and $|\eta|<0.5$ were selected and subsequently reclustered using the Cambridge/Aachen algorithm \cite{Dokshitzer:1997in}, which has been shown to provide the best performance when it comes to resolve jet substructure \cite{Dreyer:2018nbf}. The Lund plane was populated by going backwards through the clustering sequence, i.e.\ taking the jet and declustering it into two subjets $p_a$ and $p_b$ with $p_b$ having the smaller transverse momentum. It should be noted that we have deliberately not limited the jet to a certain reconstructed jet $p_{\rm T}$ bin and instead fixed the generated $\hat{p}_{\rm T}$ range in order to minimise biases induced in the final-state jet selection and to be able to compare the effect for the same parton $p_T$.

The results are seen in Fig.~\ref{fig:lund_plane}, were we show the JEWEL predictions for the full Lund plane for (a) pp collisions (top left) and (b-d) the difference of pp and central PbPb collisions when including (b) only cold nuclear matter effects (top right), (c) in addition medium effects (bottom left) and finally (d) cold nuclear matter, medium and recoil effects (bottom right). When we compare the first two figures (a) and (b), we see that the impact of initial-state modifications from nuclear PDFs on the final-state and the jet shower as visualised in the Lund plane is very small as expected. Qualitatively and with larger fluctuations, we observe fewer hard and more soft subjets. On a much larger scale, the depletion of hard subjets is seen when we include the medium (c) and recoils (d). It is remarkable that the addition of a coherent background does not alter the magnitude of the depletion, but rather provides a more balanced distribution, since at all angles now a clear increase of soft radiation is visible.       

\section{Conclusion}
\label{sec:4}

In this paper, we have made theoretical predictions based on JEWEL and PYTHIA 6.4 for inclusive jet cross sections in pp collisions, jet yields in PbPb collisions, and their ratio, the nuclear modification factor $R_{\rm AA}$, and compared these predictions to the corresponding measurements from the ALICE, CMS and ATLAS collaborations. These comparisons were made for two centrality ranges, three different jet radii, and two different centre-of-mass energies and the corresponding accessible jet transverse-momentum ranges.

We first established compatibility of the different data sets and qualitative agreement of our predictions with the pp and PbPb measurements. Here, our focus was on the scale uncertainty, which we found to be considerable in these LO calculations for absolute cross sections, but to be reduced not only in ratios of cross sections, but also when NLO plus parton shower corrections from POWHEG were employed to predict the optimal scale. The uncertainty from nuclear PDFs as estimated within both the nCTEQ15 and EPPS16 global fits was also significant and remained so also in the ratio $R_{\rm AA}$, so that it affected the determination of the medium parameters, in particular the initial temperature of the medium. A disentangling of the two effects became only possible when measurements from different centralities were compared. A reduction of the nuclear PDF uncertainty would require input from more data, in particular from pA collisions at the LHC \cite{Brandt:2014vva,Kusina:2017gkz,Guzey:2019kik,Eskola:2019bgf,Kusina:2020lyz} or eA collisions at the future EIC \cite{Aschenauer:2017oxs,Klasen:2017kwb,Klasen:2018gtb}. Nevertheless, in most cases we observed quantitatively good agreement of the central nCTEQ15 and EPPS16 predictions with the data within the experimental uncertainties.

The JEWEL predictions were found to depend significantly on the initial temperature, which was higher than expected (420 MeV) for 0-10\% than for 10-30\% central collisions (360 MeV). In contrast, the dependence on the crossover temperature, where the final-state phase transition from partons to hadrons occurs, was very small, again as expected, since parton-hadron duality is a key feature of well-defined jets. At low transverse momentum a small preference of the data for the now favoured lower value of about 160 MeV over the default value in JEWEL of 170 MeV was found.

A more detailed understanding of the jet formation and how it is affected by the medium can be obtained by studying less inclusive observables like the Lund jet plane. While the modification from free proton to nuclear PDFs in the initial state had little effect, as expected, the medium and the recoil modified the average primary Lund plane density significantly, i.e.\ due to the jet quenching, fewer hard and more soft subjets were observed, again in accordance with theoretical expectations.

While this paper was being completed, the ALICE collaboration published measurements of
inclusive jet spectra in pp and 0-10\% central PbPb collisions at $\sqrt{s_\mathrm{NN}}=5.02$~TeV
with jet radii of $R=0.2$ and 0.4 and $p_{\rm T}$-ranges of 40 and 60 to 140 GeV,
respectively~\cite{Acharya:2019jyg}. While these measurements are qualitatively similar to those at 2.76 TeV,
it will be interesting to compare our calculations in detail with these measurements in the
future.
As a further outlook, this study could be made more precise by combining NLO plus parton shower predictions in POWHEG with the medium effects in JEWEL, which would further reduce the scale uncertainty. In the long run, the connection of the initial gluon density including recombination and saturation effects with the medium model and its initial temperature should be investigated further and the relation with the alternative description within the colour-glass-condensate 
should be worked out in detail.

\acknowledgments

We thank A.\ Mazeliauskas and an anonymous referee for comments which led us to correct the scale
variation in some of our numerical results. 
This work has been supported by the Deutsche Forschungsgemeinschaft (DFG, German Research Foundation) through the Research Training Network 2149 ``Strong and weak interactions - from hadrons to dark matter''  and through Project-Id 273811115 - SFB 1225 ``ISOQUANT''. Additional support has been provided by the Bundesministerium f\"ur Bildung und Forschung (BMBF) within the ErUM-Framework (05P19PMCA1). The calculations have been performed on the high-performance computing cluster PALMA II at WWU M\"unster.

\bibliographystyle{JHEP}
\bibliography{RAA}

\providecommand{\href}[2]{#2}\begingroup\raggedright\begin{thebibliography}{10}

\bibitem{Borsanyi:2013bia}
S.~Borsanyi, Z.~Fodor, C.~Hoelbling, S.~D. Katz, S.~Krieg and K.~K. Szabo,
  \emph{{Full result for the QCD equation of state with 2+1 flavors}},
  \href{https://doi.org/10.1016/j.physletb.2014.01.007}{\emph{Phys. Lett. B}
  {\bfseries 730} (2014) 99} [\href{https://arxiv.org/abs/1309.5258}{{\ttfamily
  1309.5258}}].

\bibitem{Bazavov:2018mes}
{\scshape HotQCD} collaboration, \emph{{Chiral crossover in QCD at zero and
  non-zero chemical potentials}},
  \href{https://doi.org/10.1016/j.physletb.2019.05.013}{\emph{Phys. Lett.}
  {\bfseries B795} (2019) 15}
  [\href{https://arxiv.org/abs/1812.08235}{{\ttfamily 1812.08235}}].

\bibitem{Andronic:2018cr}
A.~Andronic, P.~Braun-Munzinger, K.~Redlich and J.~Stachel, \emph{{Decoding the
  phase structure of QCD via particle production at high energy}},
  \href{https://doi.org/10.1038/s41586-018-0491-6}{\emph{Nature} {\bfseries
  561} (2018) 321}.

\bibitem{Adcox:2001jp}
{\scshape PHENIX} collaboration, \emph{{Suppression of hadrons with large
  transverse momentum in central Au+Au collisions at $\sqrt{s_{NN}}$ =
  130-GeV}}, \href{https://doi.org/10.1103/PhysRevLett.88.022301}{\emph{Phys.
  Rev. Lett.} {\bfseries 88} (2002) 022301}
  [\href{https://arxiv.org/abs/nucl-ex/0109003}{{\ttfamily nucl-ex/0109003}}].

\bibitem{Miller:2007ri}
M.~L. Miller, K.~Reygers, S.~J. Sanders and P.~Steinberg, \emph{{Glauber
  modeling in high energy nuclear collisions}},
  \href{https://doi.org/10.1146/annurev.nucl.57.090506.123020}{\emph{Ann. Rev.
  Nucl. Part. Sci.} {\bfseries 57} (2007) 205}
  [\href{https://arxiv.org/abs/nucl-ex/0701025}{{\ttfamily nucl-ex/0701025}}].

\bibitem{Adam:2015ewa}
{\scshape ALICE} collaboration, \emph{{Measurement of jet suppression in
  central Pb-Pb collisions at $\sqrt{s_\mathrm{NN}}$ = 2.76 TeV}},
  \href{https://doi.org/10.1016/j.physletb.2015.04.039}{\emph{Phys. Lett.}
  {\bfseries B746} (2015) 1}
  [\href{https://arxiv.org/abs/1502.01689}{{\ttfamily 1502.01689}}].

\bibitem{Khachatryan:2016jfl}
{\scshape CMS} collaboration, \emph{{Measurement of inclusive jet cross
  sections in $pp$ and PbPb collisions at $\sqrt{s_{NN}}=$ 2.76 TeV}},
  \href{https://doi.org/10.1103/PhysRevC.96.015202}{\emph{Phys. Rev.}
  {\bfseries C96} (2017) 015202}
  [\href{https://arxiv.org/abs/1609.05383}{{\ttfamily 1609.05383}}].

\bibitem{Aaboud:2018twu}
{\scshape ATLAS} collaboration, \emph{{Measurement of the nuclear modification
  factor for inclusive jets in Pb+Pb collisions at $\sqrt{s_\mathrm{NN}}=5.02$
  TeV with the ATLAS detector}},
  \href{https://doi.org/10.1016/j.physletb.2018.10.076}{\emph{Phys. Lett.}
  {\bfseries B790} (2019) 108}
  [\href{https://arxiv.org/abs/1805.05635}{{\ttfamily 1805.05635}}].

\bibitem{Abelev:2013fn}
{\scshape ALICE} collaboration, \emph{{Measurement of the inclusive
  differential jet cross section in $pp$ collisions at $\sqrt{s} = 2.76$ TeV}},
  \href{https://doi.org/10.1016/j.physletb.2013.04.026}{\emph{Phys. Lett.}
  {\bfseries B722} (2013) 262}
  [\href{https://arxiv.org/abs/1301.3475}{{\ttfamily 1301.3475}}].

\bibitem{Chatrchyan:2011ua}
{\scshape CMS} collaboration, \emph{{Study of Z boson production in PbPb
  collisions at $\sqrt{s_{NN}}$ = 2.76 TeV}},
  \href{https://doi.org/10.1103/PhysRevLett.106.212301}{\emph{Phys. Rev. Lett.}
  {\bfseries 106} (2011) 212301}
  [\href{https://arxiv.org/abs/1102.5435}{{\ttfamily 1102.5435}}].

\bibitem{Chatrchyan:2012nt}
{\scshape CMS} collaboration, \emph{{Study of $W$ boson production in PbPb and
  $pp$ collisions at $\sqrt{s_{NN}}=2.76$ TeV}},
  \href{https://doi.org/10.1016/j.physletb.2012.07.025}{\emph{Phys. Lett. B}
  {\bfseries 715} (2012) 66} [\href{https://arxiv.org/abs/1205.6334}{{\ttfamily
  1205.6334}}].

\bibitem{Acharya:2017wpf}
{\scshape ALICE} collaboration, \emph{{Measurement of Z$^0$-boson production at
  large rapidities in Pb-Pb collisions at $\sqrt{s_{\rm NN}}=5.02$ TeV}},
  \href{https://doi.org/10.1016/j.physletb.2018.03.010}{\emph{Phys. Lett. B}
  {\bfseries 780} (2018) 372}
  [\href{https://arxiv.org/abs/1711.10753}{{\ttfamily 1711.10753}}].

\bibitem{Aad:2019sfe}
{\scshape ATLAS} collaboration, \emph{{Measurement of $W^\pm $ boson production
  in Pb+Pb collisions at $\sqrt{s_{\mathrm{NN}}} = 5.02~\text {Te}\text {V}$
  with the ATLAS detector}},
  \href{https://doi.org/10.1140/epjc/s10052-019-7439-3}{\emph{Eur. Phys. J. C}
  {\bfseries 79} (2019) 935}
  [\href{https://arxiv.org/abs/1907.10414}{{\ttfamily 1907.10414}}].

\bibitem{Aad:2019lan}
{\scshape ATLAS} collaboration, \emph{{$Z$ boson production in Pb+Pb collisions
  at $\sqrt{s_{\textrm{NN}}}$= 5.02 TeV measured by the ATLAS experiment}},
  \href{https://doi.org/10.1016/j.physletb.2020.135262}{\emph{Phys. Lett. B}
  {\bfseries 802} (2020) 135262}
  [\href{https://arxiv.org/abs/1910.13396}{{\ttfamily 1910.13396}}].

\bibitem{Abelev:2014dsa}
{\scshape ALICE} collaboration, \emph{{Transverse momentum dependence of
  inclusive primary charged-particle production in p-Pb collisions at
  $\sqrt{s_\mathrm{{NN}}}=5.02~\text {TeV}$}},
  \href{https://doi.org/10.1140/epjc/s10052-014-3054-5}{\emph{Eur. Phys. J.}
  {\bfseries C74} (2014) 3054}
  [\href{https://arxiv.org/abs/1405.2737}{{\ttfamily 1405.2737}}].

\bibitem{Adam:2015hoa}
{\scshape ALICE} collaboration, \emph{{Measurement of charged jet production
  cross sections and nuclear modification in p-Pb collisions at
  $\sqrt{s_\mathrm{NN}} = 5.02$ TeV}},
  \href{https://doi.org/10.1016/j.physletb.2015.07.054}{\emph{Phys. Lett.}
  {\bfseries B749} (2015) 68}
  [\href{https://arxiv.org/abs/1503.00681}{{\ttfamily 1503.00681}}].

\bibitem{Khachatryan:2016xdg}
{\scshape CMS} collaboration, \emph{{Measurement of inclusive jet production
  and nuclear modifications in pPb collisions at $\sqrt{s_{_\mathrm {NN}}} =$
  5.02 TeV}}, \href{https://doi.org/10.1140/epjc/s10052-016-4205-7}{\emph{Eur.
  Phys. J. C} {\bfseries 76} (2016) 372}
  [\href{https://arxiv.org/abs/1601.02001}{{\ttfamily 1601.02001}}].

\bibitem{Cao:2020wlm}
S.~Cao and X.-N. Wang, \emph{{Jet quenching and medium response in high-energy
  heavy-ion collisions: a review}},
  \href{https://arxiv.org/abs/2002.04028}{{\ttfamily 2002.04028}}.

\bibitem{Zapp:2008gi}
K.~Zapp, G.~Ingelman, J.~Rathsman, J.~Stachel and U.~A. Wiedemann, \emph{{{A
  Monte Carlo Model for Jet Quenching}}},
  \href{https://doi.org/10.1140/epjc/s10052-009-0941-2}{\emph{Eur. Phys. J.}
  {\bfseries C60} (2009) 617}
  [\href{https://arxiv.org/abs/0804.3568}{{\ttfamily 0804.3568}}].

\bibitem{Zapp:2012ak}
K.~C. Zapp, F.~Krauss and U.~A. Wiedemann, \emph{{A perturbative framework for
  jet quenching}}, \href{https://doi.org/10.1007/JHEP03(2013)080}{\emph{JHEP}
  {\bfseries 03} (2013) 080} [\href{https://arxiv.org/abs/1212.1599}{{\ttfamily
  1212.1599}}].

\bibitem{Zapp:2013vla}
K.~C. Zapp, \emph{{JEWEL 2.0.0: directions for use}},
  \href{https://doi.org/10.1140/epjc/s10052-014-2762-1}{\emph{Eur. Phys. J.}
  {\bfseries C74} (2014) 2762}
  [\href{https://arxiv.org/abs/1311.0048}{{\ttfamily 1311.0048}}].

\bibitem{KunnawalkamElayavalli:2017hxo}
R.~Kunnawalkam~Elayavalli and K.~C. Zapp, \emph{{Medium response in JEWEL and
  its impact on jet shape observables in heavy ion collisions}},
  \href{https://doi.org/10.1007/JHEP07(2017)141}{\emph{JHEP} {\bfseries 07}
  (2017) 141} [\href{https://arxiv.org/abs/1707.01539}{{\ttfamily
  1707.01539}}].

\bibitem{Dreyer:2018nbf}
F.~A. Dreyer, G.~P. Salam and G.~Soyez, \emph{{The Lund Jet Plane}},
  \href{https://doi.org/10.1007/JHEP12(2018)064}{\emph{JHEP} {\bfseries 12}
  (2018) 064} [\href{https://arxiv.org/abs/1807.04758}{{\ttfamily
  1807.04758}}].

\bibitem{Sjostrand:2006za}
T.~Sjostrand, S.~Mrenna and P.~Z. Skands, \emph{{PYTHIA 6.4 Physics and
  Manual}}, \href{https://doi.org/10.1088/1126-6708/2006/05/026}{\emph{JHEP}
  {\bfseries 05} (2006) 026}
  [\href{https://arxiv.org/abs/hep-ph/0603175}{{\ttfamily hep-ph/0603175}}].

\bibitem{Zapp:2013zya}
K.~C. Zapp, \emph{{Geometrical aspects of jet quenching in JEWEL}},
  \href{https://doi.org/10.1016/j.physletb.2014.06.020}{\emph{Phys. Lett. B}
  {\bfseries 735} (2014) 157}
  [\href{https://arxiv.org/abs/1312.5536}{{\ttfamily 1312.5536}}].

\bibitem{Bjorken:1982qr}
J.~D. Bjorken, \emph{{Highly Relativistic Nucleus-Nucleus Collisions: The
  Central Rapidity Region}},
  \href{https://doi.org/10.1103/PhysRevD.27.140}{\emph{Phys. Rev.} {\bfseries
  D27} (1983) 140}.

\bibitem{Eskola:2009uj}
K.~J. Eskola, H.~Paukkunen and C.~A. Salgado, \emph{{EPS09: A New Generation of
  NLO and LO Nuclear Parton Distribution Functions}},
  \href{https://doi.org/10.1088/1126-6708/2009/04/065}{\emph{JHEP} {\bfseries
  04} (2009) 065} [\href{https://arxiv.org/abs/0902.4154}{{\ttfamily
  0902.4154}}].

\bibitem{Pumplin:2002vw}
J.~Pumplin et~al., \emph{{New generation of parton distributions with
  uncertainties from global QCD analysis}},
  \href{https://doi.org/10.1088/1126-6708/2002/07/012}{\emph{JHEP} {\bfseries
  07} (2002) 012} [\href{https://arxiv.org/abs/hep-ph/0201195}{{\ttfamily
  hep-ph/0201195}}].

\bibitem{Whalley:2005nh}
M.~R. Whalley, D.~Bourilkov and R.~C. Group, \emph{{The Les Houches accord PDFs
  (LHAPDF) and LHAGLUE}},  in \emph{{HERA and the LHC: A Workshop on the
  implications of HERA for LHC physics. Proceedings, Part B}}, pp.~575--581,
  2005, \href{https://arxiv.org/abs/hep-ph/0508110}{{\ttfamily
  hep-ph/0508110}}.

\bibitem{Buckley:2014ana}
A.~Buckley, J.~Ferrando, S.~Lloyd, K.~Nordström, B.~Page, M.~Rüfenacht
  et~al., \emph{{LHAPDF6: parton density access in the LHC precision era}},
  \href{https://doi.org/10.1140/epjc/s10052-015-3318-8}{\emph{Eur. Phys. J.}
  {\bfseries C75} (2015) 132}
  [\href{https://arxiv.org/abs/1412.7420}{{\ttfamily 1412.7420}}].

\bibitem{Kovarik:2015cma}
K.~Kovarik et~al., \emph{{nCTEQ15 - Global analysis of nuclear parton
  distributions with uncertainties in the CTEQ framework}},
  \href{https://doi.org/10.1103/PhysRevD.93.085037}{\emph{Phys. Rev.}
  {\bfseries D93} (2016) 085037}
  [\href{https://arxiv.org/abs/1509.00792}{{\ttfamily 1509.00792}}].

\bibitem{Eskola:2016oht}
K.~J. Eskola, P.~Paakkinen, H.~Paukkunen and C.~A. Salgado, \emph{{EPPS16:
  Nuclear parton distributions with LHC data}},
  \href{https://doi.org/10.1140/epjc/s10052-017-4725-9}{\emph{Eur. Phys. J.}
  {\bfseries C77} (2017) 163}
  [\href{https://arxiv.org/abs/1612.05741}{{\ttfamily 1612.05741}}].

\bibitem{Dulat:2015mca}
S.~Dulat, T.-J. Hou, J.~Gao, M.~Guzzi, J.~Huston, P.~Nadolsky et~al.,
  \emph{{New parton distribution functions from a global analysis of quantum
  chromodynamics}},
  \href{https://doi.org/10.1103/PhysRevD.93.033006}{\emph{Phys. Rev. D}
  {\bfseries 93} (2016) 033006}
  [\href{https://arxiv.org/abs/1506.07443}{{\ttfamily 1506.07443}}].

\bibitem{Chetyrkin:2000yt}
K.~G. Chetyrkin, J.~H. K\"uhn and M.~Steinhauser, \emph{{RunDec: A Mathematica
  package for running and decoupling of the strong coupling and quark masses}},
  \href{https://doi.org/10.1016/S0010-4655(00)00155-7}{\emph{Comput. Phys.
  Commun.} {\bfseries 133} (2000) 43}
  [\href{https://arxiv.org/abs/hep-ph/0004189}{{\ttfamily hep-ph/0004189}}].

\bibitem{Brandt:2014vva}
M.~Brandt, M.~Klasen and F.~K\"onig, \emph{{Nuclear parton density
  modifications from low-mass lepton pair production at the LHC}},
  \href{https://doi.org/10.1016/j.nuclphysa.2014.03.024}{\emph{Nucl. Phys. A}
  {\bfseries 927} (2014) 78} [\href{https://arxiv.org/abs/1401.6817}{{\ttfamily
  1401.6817}}].

\bibitem{Kusina:2017gkz}
A.~Kusina, J.-P. Lansberg, I.~Schienbein and H.-S. Shao, \emph{{Gluon Shadowing
  in Heavy-Flavor Production at the LHC}},
  \href{https://doi.org/10.1103/PhysRevLett.121.052004}{\emph{Phys. Rev. Lett.}
  {\bfseries 121} (2018) 052004}
  [\href{https://arxiv.org/abs/1712.07024}{{\ttfamily 1712.07024}}].

\bibitem{Guzey:2019kik}
V.~Guzey and M.~Klasen, \emph{{Constraints on nuclear parton distributions from
  dijet photoproduction at the LHC}},
  \href{https://doi.org/10.1140/epjc/s10052-019-6905-2}{\emph{Eur. Phys. J. C}
  {\bfseries 79} (2019) 396}
  [\href{https://arxiv.org/abs/1902.05126}{{\ttfamily 1902.05126}}].

\bibitem{Eskola:2019bgf}
K.~J. Eskola, I.~Helenius, P.~Paakkinen and H.~Paukkunen, \emph{{A QCD analysis
  of LHCb D-meson data in p+Pb collisions}},
  \href{https://doi.org/10.1007/JHEP05(2020)037}{\emph{JHEP} {\bfseries 05}
  (2020) 037} [\href{https://arxiv.org/abs/1906.02512}{{\ttfamily
  1906.02512}}].

\bibitem{Kusina:2020lyz}
A.~Kusina et~al., \emph{{Impact of LHC vector boson production in heavy ion
  collisions on strange PDFs}},
  \href{https://doi.org/10.1140/epjc/s10052-020-08532-4}{\emph{Eur. Phys. J. C}
  {\bfseries 80} (2020) 968}
  [\href{https://arxiv.org/abs/2007.09100}{{\ttfamily 2007.09100}}].

\bibitem{Segarra:2020gtj}
E.~P. Segarra et~al., \emph{{nCTEQ15HIX -- Extending nPDF Analyses into the
  High-$x$, Low $Q^2$ Region}},
  \href{https://arxiv.org/abs/2012.11566}{{\ttfamily 2012.11566}}.

\bibitem{AbdulKhalek:2019mzd}
{\scshape NNPDF} collaboration, \emph{{Nuclear parton distributions from
  lepton-nucleus scattering and the impact of an electron-ion collider}},
  \href{https://doi.org/10.1140/epjc/s10052-019-6983-1}{\emph{Eur. Phys. J. C}
  {\bfseries 79} (2019) 471}
  [\href{https://arxiv.org/abs/1904.00018}{{\ttfamily 1904.00018}}].

\bibitem{Walt:2019slu}
M.~Walt, I.~Helenius and W.~Vogelsang, \emph{{Open-source QCD analysis of
  nuclear parton distribution functions at NLO and NNLO}},
  \href{https://doi.org/10.1103/PhysRevD.100.096015}{\emph{Phys. Rev. D}
  {\bfseries 100} (2019) 096015}
  [\href{https://arxiv.org/abs/1908.03355}{{\ttfamily 1908.03355}}].

\bibitem{Alioli:2010xa}
S.~Alioli, K.~Hamilton, P.~Nason, C.~Oleari and E.~Re, \emph{{Jet pair
  production in POWHEG}},
  \href{https://doi.org/10.1007/JHEP04(2011)081}{\emph{JHEP} {\bfseries 04}
  (2011) 081} [\href{https://arxiv.org/abs/1012.3380}{{\ttfamily 1012.3380}}].

\bibitem{Klasen:1994bj}
M.~Klasen, G.~Kramer and S.~Salesch, \emph{{Photoproduction of jets at HERA:
  Comparison of next-to-leading order calculation with ZEUS data}},
  \href{https://doi.org/10.1007/BF01579810}{\emph{Z. Phys. C} {\bfseries 68}
  (1995) 113}.

\bibitem{Klasen:1996yk}
M.~Klasen and G.~Kramer, \emph{{Large transverse momentum jet production and
  DIS distributions of the proton}},
  \href{https://doi.org/10.1016/0370-2693(96)00960-4}{\emph{Phys. Lett. B}
  {\bfseries 386} (1996) 384}
  [\href{https://arxiv.org/abs/hep-ph/9605210}{{\ttfamily hep-ph/9605210}}].

\bibitem{Loizides:2017ack}
C.~Loizides, J.~Kamin and D.~d'Enterria, \emph{{Improved Monte Carlo Glauber
  predictions at present and future nuclear colliders}},
  \href{https://doi.org/10.1103/PhysRevC.97.054910}{\emph{Phys. Rev. C}
  {\bfseries 97} (2018) 054910}
  [\href{https://arxiv.org/abs/1710.07098}{{\ttfamily 1710.07098}}].

\bibitem{Shen:2012vn}
C.~Shen and U.~Heinz, \emph{{Collision Energy Dependence of Viscous
  Hydrodynamic Flow in Relativistic Heavy-Ion Collisions}},
  \href{https://doi.org/10.1103/PhysRevC.86.049903,
  10.1103/PhysRevC.85.054902}{\emph{Phys. Rev.} {\bfseries C85} (2012) 054902}
  [\href{https://arxiv.org/abs/1202.6620}{{\ttfamily 1202.6620}}].

\bibitem{Dorau:2019ozd}
P.~Dorau, J.-B. Rose, D.~Pablos and H.~Elfner, \emph{{Jet Quenching in the
  Hadron Gas: An Exploratory Study}},
  \href{https://doi.org/10.1103/PhysRevC.101.035208}{\emph{Phys. Rev. C}
  {\bfseries 101} (2020) 035208}
  [\href{https://arxiv.org/abs/1910.07027}{{\ttfamily 1910.07027}}].

\bibitem{Cacciari:2008gp}
M.~Cacciari, G.~P. Salam and G.~Soyez, \emph{{The anti-$k_t$ jet clustering
  algorithm}}, \href{https://doi.org/10.1088/1126-6708/2008/04/063}{\emph{JHEP}
  {\bfseries 04} (2008) 063} [\href{https://arxiv.org/abs/0802.1189}{{\ttfamily
  0802.1189}}].

\bibitem{Cacciari:2011ma}
M.~Cacciari, G.~P. Salam and G.~Soyez, \emph{{FastJet User Manual}},
  \href{https://doi.org/10.1140/epjc/s10052-012-1896-2}{\emph{Eur. Phys. J.}
  {\bfseries C72} (2012) 1896}
  [\href{https://arxiv.org/abs/1111.6097}{{\ttfamily 1111.6097}}].

\bibitem{Maguire:2017ypu}
E.~Maguire, L.~Heinrich and G.~Watt, \emph{{HEPData: a repository for high
  energy physics data}},
  \href{https://doi.org/10.1088/1742-6596/898/10/102006}{\emph{J. Phys. Conf.
  Ser.} {\bfseries 898} (2017) 102006}
  [\href{https://arxiv.org/abs/1704.05473}{{\ttfamily 1704.05473}}].

\bibitem{Buckley:2010ar}
A.~Buckley, J.~Butterworth, L.~Lonnblad, D.~Grellscheid, H.~Hoeth, J.~Monk
  et~al., \emph{{Rivet user manual}},
  \href{https://doi.org/10.1016/j.cpc.2013.05.021}{\emph{Comput. Phys. Commun.}
  {\bfseries 184} (2013) 2803}
  [\href{https://arxiv.org/abs/1003.0694}{{\ttfamily 1003.0694}}].

\bibitem{Huss:2020dwe}
A.~Huss, A.~Kurkela, A.~Mazeliauskas, R.~Paatelainen, W.~van~der Schee and
  U.~A. Wiedemann, \emph{{Discovering partonic rescattering in light nucleus
  collisions}},  \href{https://arxiv.org/abs/2007.13754}{{\ttfamily
  2007.13754}}.

\bibitem{Romatschke:2017ejr}
P.~Romatschke and U.~Romatschke, \emph{{Relativistic Fluid Dynamics In and Out
  of Equilibrium}}, Cambridge Monographs on Mathematical Physics. Cambridge
  University Press, 5, 2019,
  \href{https://doi.org/10.1017/9781108651998}{10.1017/9781108651998},
  [\href{https://arxiv.org/abs/1712.05815}{{\ttfamily 1712.05815}}].

\bibitem{Busza:2018rrf}
W.~Busza, K.~Rajagopal and W.~van~der Schee, \emph{{Heavy Ion Collisions: The
  Big Picture, and the Big Questions}},
  \href{https://doi.org/10.1146/annurev-nucl-101917-020852}{\emph{Ann. Rev.
  Nucl. Part. Sci.} {\bfseries 68} (2018) 339}
  [\href{https://arxiv.org/abs/1802.04801}{{\ttfamily 1802.04801}}].

\bibitem{Albacete:2014fwa}
J.~L. Albacete and C.~Marquet, \emph{{Gluon saturation and initial conditions
  for relativistic heavy ion collisions}},
  \href{https://doi.org/10.1016/j.ppnp.2014.01.004}{\emph{Prog. Part. Nucl.
  Phys.} {\bfseries 76} (2014) 1}
  [\href{https://arxiv.org/abs/1401.4866}{{\ttfamily 1401.4866}}].

\bibitem{Rezaeian:2012ji}
A.~H. Rezaeian, M.~Siddikov, M.~Van~de Klundert and R.~Venugopalan,
  \emph{{Analysis of combined HERA data in the Impact-Parameter dependent
  Saturation model}},
  \href{https://doi.org/10.1103/PhysRevD.87.034002}{\emph{Phys. Rev. D}
  {\bfseries 87} (2013) 034002}
  [\href{https://arxiv.org/abs/1212.2974}{{\ttfamily 1212.2974}}].

\bibitem{Golec-Biernat:2017lfv}
K.~Golec-Biernat and S.~Sapeta, \emph{{Saturation model of DIS : an update}},
  \href{https://doi.org/10.1007/JHEP03(2018)102}{\emph{JHEP} {\bfseries 03}
  (2018) 102} [\href{https://arxiv.org/abs/1711.11360}{{\ttfamily
  1711.11360}}].

\bibitem{Levin:2010dw}
E.~Levin and A.~H. Rezaeian, \emph{{Gluon saturation and inclusive hadron
  production at LHC}},
  \href{https://doi.org/10.1103/PhysRevD.82.014022}{\emph{Phys. Rev. D}
  {\bfseries 82} (2010) 014022}
  [\href{https://arxiv.org/abs/1005.0631}{{\ttfamily 1005.0631}}].

\bibitem{Dusling:2012cg}
K.~Dusling and R.~Venugopalan, \emph{{Evidence for BFKL and saturation dynamics
  from dihadron spectra at the LHC}},
  \href{https://doi.org/10.1103/PhysRevD.87.051502}{\emph{Phys. Rev. D}
  {\bfseries 87} (2013) 051502}
  [\href{https://arxiv.org/abs/1210.3890}{{\ttfamily 1210.3890}}].

\bibitem{Mantysaari:2018nng}
H.~M\"antysaari and P.~Zurita, \emph{{In depth analysis of the combined HERA
  data in the dipole models with and without saturation}},
  \href{https://doi.org/10.1103/PhysRevD.98.036002}{\emph{Phys. Rev. D}
  {\bfseries 98} (2018) 036002}
  [\href{https://arxiv.org/abs/1804.05311}{{\ttfamily 1804.05311}}].

\bibitem{Berges:2020fwq}
J.~Berges, M.~P. Heller, A.~Mazeliauskas and R.~Venugopalan,
  \emph{{Thermalization in QCD: theoretical approaches, phenomenological
  applications, and interdisciplinary connections}},
  \href{https://arxiv.org/abs/2005.12299}{{\ttfamily 2005.12299}}.

\bibitem{Klasen:2013mga}
M.~Klasen, C.~Klein-B\"osing, F.~K\"onig and J.~Wessels, \emph{{How robust is a
  thermal photon interpretation of the ALICE low-$p_T$ data?}},
  \href{https://doi.org/10.1007/JHEP10(2013)119}{\emph{JHEP} {\bfseries 10}
  (2013) 119} [\href{https://arxiv.org/abs/1307.7034}{{\ttfamily 1307.7034}}].

\bibitem{Adam:2015lda}
{\scshape ALICE} collaboration, \emph{{Direct photon production in Pb-Pb
  collisions at $\sqrt{s_{\mathrm{NN}}} =$ 2.76 TeV}},
  \href{https://doi.org/10.1016/j.physletb.2016.01.020}{\emph{Phys. Lett.}
  {\bfseries B754} (2016) 235}
  [\href{https://arxiv.org/abs/1509.07324}{{\ttfamily 1509.07324}}].

\bibitem{Azimov:1984np}
Y.~I. Azimov, Y.~L. Dokshitzer, V.~A. Khoze and S.~Troyan, \emph{{Similarity of
  Parton and Hadron Spectra in QCD Jets}},
  \href{https://doi.org/10.1007/BF01642482}{\emph{Z. Phys. C} {\bfseries 27}
  (1985) 65}.

\bibitem{Klasen:1997tj}
M.~Klasen and G.~Kramer, \emph{{Jet shapes in $e p$ and $p \bar{p}$ collisions
  in NLO QCD}}, \href{https://doi.org/10.1103/PhysRevD.56.2702}{\emph{Phys.
  Rev. D} {\bfseries 56} (1997) 2702}
  [\href{https://arxiv.org/abs/hep-ph/9701247}{{\ttfamily hep-ph/9701247}}].

\bibitem{Biswas:2019jkr}
R.~Biswas, S.~Choudhury, S.~K. Prasad and S.~Das, \emph{{Study of jet-medium
  interactions using jet shape observables in heavy ion collisions at LHC
  energies with JEWEL}},
  \href{https://doi.org/10.1088/1361-6471/ab2e69}{\emph{J. Phys. G} {\bfseries
  46} (2019) 095004} [\href{https://arxiv.org/abs/1906.05513}{{\ttfamily
  1906.05513}}].

\bibitem{Andersson:1988gp}
B.~Andersson, G.~Gustafson, L.~Lonnblad and U.~Pettersson, \emph{{Coherence
  Effects in Deep Inelastic Scattering}},
  \href{https://doi.org/10.1007/BF01550942}{\emph{Z. Phys. C} {\bfseries 43}
  (1989) 625}.

\bibitem{Dasgupta:2013ihk}
M.~Dasgupta, A.~Fregoso, S.~Marzani and G.~P. Salam, \emph{{Towards an
  understanding of jet substructure}},
  \href{https://doi.org/10.1007/JHEP09(2013)029}{\emph{JHEP} {\bfseries 09}
  (2013) 029} [\href{https://arxiv.org/abs/1307.0007}{{\ttfamily 1307.0007}}].

\bibitem{Andrews:2018jcm}
H.~A. Andrews et~al., \emph{{Novel tools and observables for jet physics in
  heavy-ion collisions}},
  \href{https://doi.org/10.1088/1361-6471/ab7cbc}{\emph{J. Phys. G} {\bfseries
  47} (2020) 065102} [\href{https://arxiv.org/abs/1808.03689}{{\ttfamily
  1808.03689}}].

\bibitem{Dokshitzer:1997in}
Y.~L. Dokshitzer, G.~Leder, S.~Moretti and B.~Webber, \emph{{Better jet
  clustering algorithms}},
  \href{https://doi.org/10.1088/1126-6708/1997/08/001}{\emph{JHEP} {\bfseries
  08} (1997) 001} [\href{https://arxiv.org/abs/hep-ph/9707323}{{\ttfamily
  hep-ph/9707323}}].

\bibitem{Aschenauer:2017oxs}
E.~Aschenauer, S.~Fazio, M.~Lamont, H.~Paukkunen and P.~Zurita, \emph{{Nuclear
  Structure Functions at a Future Electron-Ion Collider}},
  \href{https://doi.org/10.1103/PhysRevD.96.114005}{\emph{Phys. Rev. D}
  {\bfseries 96} (2017) 114005}
  [\href{https://arxiv.org/abs/1708.05654}{{\ttfamily 1708.05654}}].

\bibitem{Klasen:2017kwb}
M.~Klasen, K.~Kovarik and J.~Potthoff, \emph{{Nuclear parton density functions
  from jet production in DIS at an EIC}},
  \href{https://doi.org/10.1103/PhysRevD.95.094013}{\emph{Phys. Rev. D}
  {\bfseries 95} (2017) 094013}
  [\href{https://arxiv.org/abs/1703.02864}{{\ttfamily 1703.02864}}].

\bibitem{Klasen:2018gtb}
M.~Klasen and K.~Kova\v{r}\'\i{}k, \emph{{Nuclear parton density functions from
  dijet photoproduction at the EIC}},
  \href{https://doi.org/10.1103/PhysRevD.97.114013}{\emph{Phys. Rev. D}
  {\bfseries 97} (2018) 114013}
  [\href{https://arxiv.org/abs/1803.10985}{{\ttfamily 1803.10985}}].

\bibitem{Acharya:2019jyg}
{\scshape ALICE} collaboration, \emph{{Measurements of inclusive jet spectra in
  pp and central Pb-Pb collisions at $\sqrt{s_{\rm{NN}}}$ = 5.02 TeV}},
  \href{https://doi.org/10.1103/PhysRevC.101.034911}{\emph{Phys. Rev. C}
  {\bfseries 101} (2020) 034911}
  [\href{https://arxiv.org/abs/1909.09718}{{\ttfamily 1909.09718}}].

\end{thebibliography}\endgroup

\end{document}